\documentclass[10pt,journal,epsfig]{IEEEtran}

\usepackage{graphicx}
\usepackage{amssymb}
\usepackage{cite}
\usepackage{amsmath}
\usepackage{multirow}
\usepackage{subfig}
\usepackage{stmaryrd}
\usepackage{comment}
\usepackage{epstopdf}
\usepackage{color}
\usepackage{algorithm,algpseudocode}
\graphicspath{{figures/}}

\newcommand{\bm}{\boldsymbol}

\newtheorem{proposition}{Proposition}
\newtheorem{theorem}{Theorem}
\newtheorem{remark}{Remark}

\newtheorem{definition}{Definition}

\begin{document}

\title{Near/Far-Field Channel Estimation For Terahertz Systems With ELAAs:
A Block-Sparse-Aware Approach}
\author{Hongwei Wang, Jun Fang, Huiping
Duan, Hongbin Li,~\IEEEmembership{Fellow,~IEEE}
\thanks{Hongwei Wang, and Jun Fang are with the National Key Laboratory
of Wireless Communications, University of Electronic Science and
Technology of China, Chengdu 611731, China, Email:
JunFang@uestc.edu.cn}
\thanks{Huiping Duan is with the School of Information and Communications Engineering,
University of Electronic Science and Technology of China, Chengdu
611731, China, Email: huipingduan@uestc.edu.cn}
\thanks{Hongbin Li is
with the Department of Electrical and Computer Engineering,
Stevens Institute of Technology, Hoboken, NJ 07030, USA, E-mail:
Hongbin.Li@stevens.edu} }

\maketitle

\begin{abstract}
Millimeter wave/Terahertz (mmWave/THz) communication with
extremely large-scale antenna arrays (ELAAs) offers a promising
solution to meet the escalating demand for high data rates in
next-generation communications. A large array aperture, along with
the ever increasing carrier frequency over the mmWave/THz bands,
leads to a large Rayleigh distance. As a result, the traditional
planar-wave assumption may not hold valid for mmWave/THz systems
featuring ELAAs. In this paper, we consider the problem of
hybrid near/far-field channel estimation by taking spherical
wave propagation into account. By analyzing the coherence
properties of any two near-field steering vectors, we prove that
the hybrid near/far-field channel admits a block-sparse
representation on a specially designed unitary matrix.
Specifically, the percentage of nonzero elements of such a
block-sparse representation is in the order of
$1/\sqrt{N}$, which tends to zero as the number of antennas, $N$,
grows. Such a block-sparse representation allows to convert
channel estimation into a block-sparse signal recovery problem.
Simulation results are provided to verify our theoretical results
and illustrate the performance of the proposed channel estimation
approach in comparison with existing state-of-the-art methods.
\end{abstract}



\section{Introduction}
Millimeter wave (mmWave) and extremely large-scale antenna arrays (ELAAs)
are two key enabling technologies for the next-generation mobile
communications to enhance the system capacity and
coverage~\cite{YuanZhao20,JiHan21,LiuWang23}. With the
increase of the number of antennas at the base stations (BS) and
the decrease of the signal wavelength, the Rayleigh distance of
the antenna array, which is considered as a boundary between the radiating near-field region and the far-field
region, may extend to several dozen meters. Consequently, the user
equipment (UE) may be located in either the near-field or the
far-field region. This renders the well-known plane wave
assumption, valid only for the far-field region, incompetent for correctly characterizing the hybrid near/far-field
channel.

Channel state information (CSI) acquisition is a prerequisite for
optimizing wireless communication systems. Accurate CSI is vital
to realize the full potential of ELAAs. The CSI can be obtained via the well-known
least squares (LS) method~\cite{Chang06,Karami07,Lin08}, provided
that the length of the pilot sequence is larger than the dimension
of the channel (typically equivalent to the number of antennas at
the BS). However, in scenarios with ELAAs, the traditional LS based channel estimation method
requires a large number of pilot symbols, making it impractical
for practical systems.


To alleviate the excessive training overhead issue, the inherent
sparse structure of mmWave/THz channels should be exploited. Real-world
measurements in dense-urban propagation environments reveal that
mmWave/THz channels exhibit a sparse structure in the angular
domain. By utilizing this property, the well-known OMP algorithm
was employed~\cite{LeeGil16,GurbuzYapici18,KimGil19} to estimate
the channel, which is sparsely represented via the standard Fourier
transform. In addition, the sparse Bayesian
learning~\cite{SrivastavaMishra18,WuMa22} as well as the message
passing algorithms~\cite{HuangLiu18,BelliliSohrabi19,MyersHeath19}
were utilized to further improve the channel estimation accuracy.
Furthermore, to mitigate the performance degradation caused by
grid mismatch, the atomic norm minimization based gridless
compressive sensing algorithm~\cite{TsaiZheng18} and an iterative
reweighted ``super-resolution'' compressive sensing
approach~\cite{HuDai18} were developed. In addition, the low-rank property of
the channel has been considered as another important structure
for reducing the training overhead.
In~\cite{ZhouFang17}, the channel was represented by a tensor and
then estimated via a CP-decomposition-based method. The low-rank
and the sparsity were simultaneously leveraged to devise the
channel estimation method~\cite{LiFang17}.

The aforementioned channel estimation methods, however, cannot be
directly applied to hybrid near/far-field channels. This is
because, unlike the far-field channel model, the hybrid
near/far-field channel is characterized by two parameters, namely,
the spatial angle and the distance. The near-field channel
estimation was studied in~\cite{CuiDai22}, where a sparse
representation of the near-field channel was investigated in the
two-dimensional angular-range domain, also referred to as the
polar domain. This sparsity in the polar domain allows to convert
near-field channel estimation into a compressed sensing problem.
However, joint sampling over this two-dimensional domain results
in a sparsifying dictionary with a large number of atoms, leading to an increased computational complexity. Also, due to
the correlation among near-field steering vectors, the dictionary
exhibits an unsatisfactory restricted isometry property (RIP) that has
a detrimental effect on the channel estimation performance. An
extension of the research~\cite{CuiDai22} was presented
in~\cite{WeiDai21}, which explored the co-existence of near-field and
far-field channels. The far-field components and the
near-field components were estimated separately, leveraging prior
information on the number of paths for each channel type.
Nevertheless, such prior knowledge is typically unavailable in
real-world scenarios. Moreover, based on the polar-domain channel
modeling approach, deep learning techniques were developed for
near-field channel estimation~\cite{LeeJu22,ZhangWang23,LeiZhang23}.


In addition to the polar-domain channel model, a hybrid
spherical- and planar-wave channel model (HSPM) was
proposed in~\cite{ChenYan21,YangChen23,TarboushAli24} via a
sub-array partitioning scheme. Specifically, the ELAA is divided
into several virtual sub-arrays, and the number of antenna of each
sub-array is small enough such that the well-known planar-wave
assumption is valid. In the HSPM framework, the channel
estimation generally involves two steps~\cite{TarboushAli24}. In
the first step, the channel of a reference sub-array is estimated
based on the far-field channel model. Leveraging the channel
parameters obtained in the first step, the hybrid near/far
channel can then be more efficiently estimated in the second
phase. In addition, another work~\cite{ChenYan21} developed a deep
convolutional-neural-network to estimate the channel parameters
associated with the reference sub-array, including the channel
gain, the spatial angle as well as the distance parameter of each
path. Subsequently, utilizing geometric relationships, the
entire channel vector can be estimated. Although the HSPM takes
both the spherical wave and planar wave into consideration, there
is still a lack of sufficient investigation to exploit the
inherent channel sparsity of the hybrid near/far-field channel.


In this paper, we provide an in-depth study of the coherence between
any two near-field steering vectors. Based on the derived results,
we show that as the number of antennas $N$ increases, hybrid near/far-field
channels exhibit a block-sparse representation on a specially
designed unitary matrix. Specifically, our analysis indicates that the
percentage of nonzero entries in the block-sparse vector is
at the order of $1/\sqrt{N}$, which tends to zero as $N$ grows.
This property allows to exploit the block-sparse structure
inherent in near-field channels. Consequently, hybrid
near/far-field channel estimation is converted into a
block-sparse signal recovery problem that can be efficiently
solved by many existing compressed sensing methods. Compared with
the polar-domain representation, our new representation
leads to a well-conditioned measurement matrix that is more amiable for
compressive sensing, which helps achieve a performance improvement
as well as an improved computational efficiency.

The current work is an extension of our prior conference
paper~\cite{WangFang23}. In the current work, we provide a more
comprehensive analysis of the coherence of near-field steering
vectors, some of which were not considered/included in our prior
work. In addition, we provide a theoretical analysis of the
block-RIP condition of the constructed measurement matrix, which
serves as a basis for analyzing the sample complexity of the
compressed sensing methods.

The rest of the paper is organized as follows. Section~\ref{2}
presents the system model and the hybrid near/far-field channel
model. In Section~\ref{3}, we introduce a new unitary matrix for hybrid near/far-field channel representation and show that the
near/far channel admits a block-sparse representation on the
specially constructed unitary matrix. Section~\ref{4} develops a channel
estimation approach based on our block-spare channel representation and analyzes the
block-RIP condition for the compressed sensing problem. Simulation
results are provided in Section~\ref{5}, followed by concluding
remarks in Section~\ref{6}.

\section{System Model and Problem Formulation}
\label{2}
\subsection{System Model}
We consider a downlink communication scenario, where a BS equipped
with an extremely large-scale uniform linear array (ULA) serves
multiple single-antenna users. Note that for the downlink channel estimation
problem, each user estimates its own channel based on its
received signal. Therefore, in the following, we only focus on the
channel estimation of a single user. In the training stage,
the BS sends a pilot signal $s_t = 1$ to the receiver.
The received signal at the $t$th time instant at the user can be expressed as
\begin{align}
y_t &= \bm h^H\bm f_ts_t +n_t \notag\\
&= \bm h^H\bm f_t +n_t
\end{align}
where $\bm h\in\mathbb{C}^{N\times 1}$ denotes the channel from the
BS to the user, $\bm f_t\in \mathbb{C}^{N\times 1}$ is the precoding vector, and $n_t \sim \mathbb{CN}(0,\sigma^2)$ is the additive complex
Gaussian noise. Define $\bm y \triangleq [y_1\ \cdots\ y_T]^H$,
$\bm F \triangleq [\bm f_1\ \cdots \ \bm f_T]^H$, and $\bm n
\triangleq [n_1\ \cdots\ n_T]^H$. We can express the received
signal as
\begin{align}
\bm y = \bm F \bm h + \bm n
\label{y}
\end{align}
The objective of this work is to estimate $\bm h$ by leveraging
the received noisy observations $\bm y$.

\subsection{Channel Model}
Theoretically, $\bm h$ can be estimated via a least squares (LS)
method provided that the number of observations $T$ is no less
than $N$. Nevertheless, $N$ is in general a large number for
extremely large-scale antennas. Hence the LS-based channel
estimation method may incur a prohibitively high training
overhead. Therefore, the inherent structure of $\bm h$ should be
exploited to reduce the training overhead.

According to the electromagnetic theory, the electromagnetic field
can be divided into three regions, i.e., the reactive near-field
region, the radiating near-field region, and the far-field region.
The Fresnel distance, denoted by $F_r$, is used to characterize the
boundary between the reactive near-field and the radiating
near-field, while the Rayleigh distance, denoted by $R$, is used
to characterize the boundary between the near field and far field:
\begin{align}
F_r\triangleq  0.62\sqrt{{D^3}/{\lambda}},\quad  R \triangleq {2D^2}/{\lambda} \label{Rayleigh-distance}
\end{align}
where $D $ is the array aperture and $\lambda$ is the wavelength.

For conventional massive MIMO systems with a moderate number of
antennas, the Rayleigh distance is usually small such that users
are always located in the far-field region of the BS. For example,
for a system with $8$ antennas and operating at $3.5$GHz, the
corresponding Rayleigh distance is about $2.1$m. Under such a
circumstance, the planar wave assumption usually holds valid,
which leads to the classical far-filed channel model
\begin{align}
\bm h  = \sum_{l=1}^L \tilde {g}_l e^{-j\frac{2\pi}{\lambda}r_l}
\bm a(\theta_l)\triangleq \sum_{l=1}^L {g}_l\bm a(\theta_l)
\label{far_channel}
\end{align}
where $L$ is the number of signal paths, ${g}_l\triangleq \tilde
{g}_le^{-j\frac{2\pi}{\lambda}r_l}$ with $\tilde {g}_l$ and $r_l$
denoting the channel gain and the distance between the BS and the
scatterer/user respectively, $\theta_l$ is the angle-of-departure
(AoD) of the $l$th path, and $\bm a(\theta_l)$ denotes a far-field
steering vector given as
\begin{align}
\bm a(\theta)\triangleq &\frac{1}{\sqrt{N}}\left[1\ \cdots \ e^{j\frac{2\pi}{\lambda}(n-1)d\sin(\theta)} \right.\notag \\
&\qquad\qquad\qquad\quad\left.\cdots\ e^{j\frac{2\pi}{\lambda}(N-1)d\sin(\theta)}\right]^T
\label{far_steering}
\end{align}
where $d$ denotes the distance between neighboring antennas, and $\lambda$ is the wavelength of the carrier signal.


\begin{figure}[!t]
\centering {\includegraphics[width=0.45\textwidth]{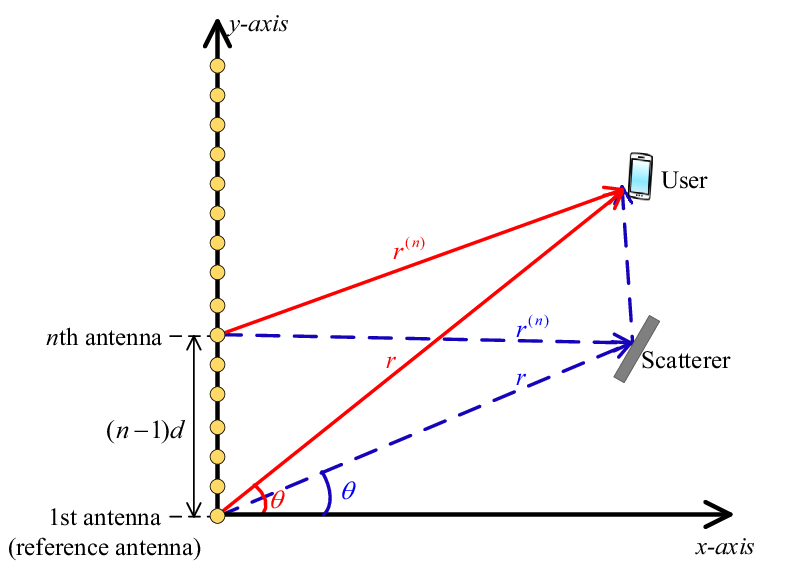} }
\caption{Schematic diagram of $r^{(n)}$} \label{f0}
\end{figure}


For extremely large-scale massive MIMO systems, hundreds or even
thousands of antennas may be deployed at the BS.
Consider a mmWave system operating at $100$GHz and equipped with
$256$ antennas. It can be calculated that the Fresnel distance is
around $2.7$m, and the Rayleigh distance is approximately $97.5$m.
From this example, we see that it is safely to assume that:
\begin{itemize}
\item[A1] The user/scatterer is located beyond the Fresnel distance, i.e., $r_l>F_r$.
\end{itemize}
Note that such an assumption is also adopted in other near-field
works, e.g.,~\cite{CuiDai22,LiuOuyang24}.
On the other hand, the user/scatterer could be very likely located
within the Rayleigh distance, i.e., $r_l < R$. To characterize the near-field channel, we adopt the following
spherical wave channel model:
\begin{align}
\bm h = \sum_{l=0}^{L-1}\tilde {g}_le^{-j\frac{2\pi}{\lambda}r_l}
\bm a(\theta_l,r_l)\triangleq \sum_{l=0}^{L-1}{g}_l\bm a(\theta_l,r_l)
\label{channel_model}
\end{align}
where $l=0$ denotes the line-of-sight (LOS) path; $l = 1,\cdots,
L-1$ represents the NLOS paths; $r_l$ denotes the distance between
the first antenna at the BS (which is considered as the reference
antenna) and the user/scatterer associated with the $l$th path;
and $\bm a(\theta,r)$ denotes a near-field steering vector which
depends not only on the angle but also on the distance.
Specifically, $\bm a(\theta,r)$ has the following expression:
\begin{align}
\bm a(\theta ,r )\triangleq
\frac{1}{\sqrt{N}}\left[e^{-j\frac{2\pi}{\lambda}(r^{(1)}-r)}\
\cdots \ e^{-j\frac{2\pi}{\lambda} (r^{(N)}-r)}\right]^T
\label{sv_nf}
\end{align}
where
$r^{(n)}$ denotes the distance between the $n$th antenna and the
scatterer/user. The schematic diagram of $r^{(n)}$ and $r$ is
plotted in Fig.~\ref{f0}, from which we can obtain the
relationship between $r^{(n)}$ and $r$ as
\begin{align}
r^{(n)} &= \sqrt{r^2 + (n-1)^2d^2 - 2r(n-1)d\sin(\theta)}\notag\\
& = r\sqrt{1+\Delta}
\label{r_nl_true}
\end{align}
where $\Delta$ is given by
\begin{align}
\Delta &\triangleq \frac{(n-1)^2d^2}{r^2}-\frac{2(n-1)d\sin(\theta)}{r}
\end{align}
In this work,~\eqref{channel_model} is referred to as the \emph{hybrid near/far-field channel model}
since it can simultaneously characterize both near-field channel and far-field channels.
Also, this general model is capable of characterizing a hybrid scenario, where the multi-path channel includes both near-field and far-field components.

The expression of $r^{(n)}$ in (\ref{r_nl_true}) is
intractable. To derive an amiable form of $r^{(n)}$, by
resorting to the Taylor expansion, one can obtain the following
widely adopted approximation of $r^{(n)}$~\cite{CuiDai22}, i.e.,
\begin{align}
r^{(n)}{\approx} -(n-1)d\sin(\theta ) + r + \frac{(n-1)^2d^2\cos^2(\theta)}{2r}
\label{r_nl}
\end{align}
It should be noted that $r^{(1)}\equiv r$ since the first antenna
is treated as a reference antenna.

Substituting (\ref{r_nl}) into~\eqref{sv_nf} leads to the following decomposition
\begin{align}
\bm a(\theta,r) \approx \bm a(\theta)\circ \bm b(\mu(r,\theta))
\label{app_sv}
\end{align}
where $\circ$ represents the Hadamard product, $\bm a(\theta)$ is
the far-field steering vector given by~\eqref{far_steering}
and $\bm b(\mu(r,\theta))$ is a unit-modulus vector given by
\begin{align}
\bm b(\mu(r,\theta))\triangleq &
\left[1\phantom{0}\cdots\phantom{0}
e^{-j\frac{2\pi}{\lambda}\big(\frac{(n-1)^2d^2}{2\mu(r,\theta)}\big)}\phantom{0}
\right.\notag \\
&\qquad\qquad\qquad\left.\cdots\phantom{0}
e^{-j\frac{2\pi}{\lambda}\big(\frac{(N-1)^2d^2}{2\mu(r,\theta)}\big)}\right]^T
\end{align}
in which $\mu(r,\theta) \triangleq r/\cos^2(\theta)$, referred to
as the {\emph{``effective distance''}}, is a parameter determined by both $r$
and $\theta$.
Here $\mu$ is referred to as the effective distance because $\mu$, which takes into
account the effect of the angle $\theta$, is more precise in
characterizing the channel phase response of the near field. Specifically, when taking the effect of the angle
into account, the effective Rayleigh distance should be defined as
~\cite[Eq.~(43)]{CuiDai24}
\begin{align}
R_{\text{eff}} = \frac{2D^2\cos^2(\theta)}{\lambda}=R
\cos^2(\theta)
\end{align}
Therefore, when the
effective distance is set larger than $R$, i.e., $\mu(r,\theta)
> R$, the corresponding physical distance $r=\mu(r,\theta)\cos^2(\theta)$, is greater than the
effective Rayleigh distance.



\section{Block-Sparsity Representation and Analysis}
\label{3}

Owing to the sparsity in the angular domain, the far-field channel
can be sparsely represented by the spatial Fourier transform
matrix~\cite{HeathGonzalez16}. For near-field channels, since it
is dependent not only on the angle but also on the distance, a
polar-domain sparse representation was introduced in~\cite{CuiDai22}, in which
a polar-domain transform matrix is constructed by simultaneously
sampling different angles and distances. Such a polar-domain
channel representation, however, contains a significant number of
atoms. In addition, these atoms are not orthogonal to each other,
which leads to an ill-conditioned sensing matrix with a relatively
large mutual coherence.

To address this issue, in this paper, we attempt to find a new
sparse representation for hybrid near/far-field channels that
facilitates channel estimation for extremely large-scale massive
MIMO systems. Inspired by~\eqref{app_sv}, we define a new
unitary matrix
\begin{align}
\bm D_{\mu} = \text{diag}\big(\bm b(\mu)\big)\bm D
\end{align}
where $\mu$ is a pre-specified effective distance and $\bm D\in
\mathbb{C}^{N\times N}$ is a discrete Fourier transform (DFT)
matrix with its $n$th column given by $\bm a(\theta_n)$ with
$\sin(\theta_n) = (2n-N-1)/N\ (n = 1,\cdots,N)$. Note that the
$n$th column of $\bm D_{\mu}$ is equal to
$\boldsymbol{b}(\mu)\circ\boldsymbol{a}(\theta_n)$, which is an
approximation of a near-field steering vector $\bm a(r,\theta_n)$ with $r$ being $\mu \cos^2(\theta_n)$.

Since $\bm b(\mu)$ is a constant modulus vector, it can be easily
verified that $\bm D_{\mu}$ is a unitary matrix
\begin{align}
\bm D_{\mu}^H \bm D_{\mu} = \bm D_{\mu}\bm D_{\mu}^H = \bm I
\end{align}
Therefore, $\bm h$ can be uniquely represented by
\begin{align}
\bm h = \bm D_{\mu} \bm \beta_{\mu}
\label{h}
\end{align}
where $\bm \beta_{\mu}$, the sparse the channel vector
on the dictionary $\bm D_{\mu}$, is given by
\begin{align}
\bm \beta_{\mu} = \sum_{l=0}^{L-1} g_l  \bm D_{\mu}^H \bm a(\theta_l,r_l)
\triangleq \sum_{l=0}^{L-1} g_l\bm \alpha_l
\label{cor1}
\end{align}
with $\bm \alpha_l\triangleq \bm
D_{\mu}^H \bm a(\theta_l,r_l)$.

We, in the following, analyze the inherent structure of $\bm
\beta_{\mu}$. Specifically, we show that, when $N$ is
sufficiently large, $\bm \beta_{\mu}$ exhibits a block-sparse
structure that can be utilized for channel estimation. Due to
sparse scattering characteristics, the number of mmWave/THz
propagation paths is usually small. To facilitate our analysis, we
first consider a simple scenario where there is only an LOS path
between the BS and the user. Based on the analysis of this simple
case, the extension of our result to the general multi-path scenario is straightforward. In the presence of a
single LOS path, $\bm \beta_{\mu}$ can be simplified as:
\begin{align}
\bm \beta_{\mu} = g_0 \bm \alpha_0
\end{align}
where $\bm \alpha_0 = \bm D_{\mu}^H \bm a(\theta_0,r_0)$.
The structure of $\bm \beta_{\mu}$ is now reduced to the structure of $\bm \alpha_0$. Note that each
entry of $\bm \alpha_0$ is an inner product of two near-field
steering vectors. Therefore to check whether $\bm \alpha_0$
exhibits a certain sparsity structure, we need to study the
coherence of two near-field steering vectors.

\begin{remark}
It should be noted that the far-field steering vector is a special case of the near-field steering vector with an infinite distance. Therefore, the subsequent results, which focus on the near-field steering vector, can be applied to the hybrid near/far-field scenarios.
\end{remark}

\subsection{Coherence of Near-Field Steering Vectors}
From the relationship $\bm \alpha_0 = \bm D_{\mu}^H \bm a(\theta_0,r_0)$, we know
that the $m$th element of $\bm \alpha_0$ is given by
\begin{align}
\bm \alpha_0(m) & = \big(\bm a(\theta_m)\circ \bm b(\mu)\big)^H \bm  a(\theta_0,r_0)\notag\\
& \overset{(a)}{=}  \big(\bm a(\theta_m)\circ \bm b(\mu)\big)^H \big(\bm  a(\theta_0) \circ \bm b(\mu_0) \big)
\label{cor2}
\end{align}
where $\mu_0 \triangleq r_0/\cos^2(\theta_0)$, and in $(a)$ we utilize the relationship in~\eqref{app_sv}. Here we assume that~\eqref{app_sv} holds strictly to ease our presentation. The approximation error can be treated as noise such that it can be absorbed into the noise term.

Define
\begin{align}
a &\triangleq \frac{2\pi d}{\lambda}\left(\sin(\theta_m)- \sin(\theta_0)\right),\label{a} \\
b &\triangleq \frac{\pi d^2}{\lambda} \left(\frac{1}{\mu_{0}} - \frac{1}{\mu}\right), \\
\Lambda_n & \triangleq (n-1)a + (n-1)^2 b, \label{Lambda-def}
\end{align}
It can be readily verified that $\bm \alpha_0(m)$ can be expressed as
\begin{align}
\bm \alpha_0(m) \triangleq f(a,b) = \frac{1}{N}\sum_{n=1}^Ne^{-j\Lambda_n}
\end{align}
Therefore, the magnitude of $f(a,b)$, denoted by $M(a,b)$, is given by
\begin{align}
M(a,b) \triangleq \left|f(a,b)\right| =\left|\frac{1}{N}\sum_{n=1}^Ne^{-j  \Lambda_n}\right|
\label{def_f}
\end{align}

Concerning the value of $M(a,b) $, we have the following results.
\begin{proposition}
\label{prop1_upt}
When $b = 0$, i.e., {\color{blue}{$\mu_0=\mu$,}}
$M(a,b) $ is given by
\begin{align}
M(a,b)=\frac{1}{N}\left|\frac{1-e^{-jNa}}{1-e^{-ja}}\right|
\end{align}
When $b \ne 0$, we have
\begin{align}
M(a,b)\approx\frac{1}{N\sqrt{|b|}} \sqrt{f_1^2(a,b) +
f_2^2(a,b)} \label{eqn11}
\end{align}
where $f_1(a,b)$ and $f_2(a,b)$ are, respectively, given as
\begin{align}
f_1(a,b) &\triangleq C\left((N-1)\sqrt{b} + \frac{\tilde a}{2\sqrt{b}}\right)
- C\left(\frac{\tilde a}{2\sqrt{b}}\right)\label{f_1}\\
f_2(a,b) &\triangleq S\left((N-1)\sqrt{b} + \frac{\tilde a}{2\sqrt{b}}\right)
- S\left(\frac{\tilde a}{2\sqrt{b}}\right)\label{f_2}\\
\tilde a &\triangleq \text{mod}(a + \pi,2\pi) - \pi\label{a_tilde}
\end{align}
with $C(\varsigma)\triangleq \int_{0}^\varsigma \cos(t^2)dt$ and
$S(\varsigma)\triangleq \int_{0}^\varsigma \sin(t^2)dt$ being,
respectively, the Fresnel integrals.
\end{proposition}
\begin{IEEEproof}
See Appendix~\ref{prop1_upt_proof}.
\end{IEEEproof}

Note that the coherence of near-field steering vectors was also studied in~\cite{CuiDai22}. Nevertheless, the work~\cite{CuiDai22} only considered the special case of $a=0$, i.e., two near-field steering vectors share the same angular parameter. Proposition~\ref{prop1_upt} generalizes the result of~\cite{CuiDai22} by considering an arbitrary value of $a$.

\begin{figure}[!t]
\centering {\includegraphics[width=0.35\textwidth]{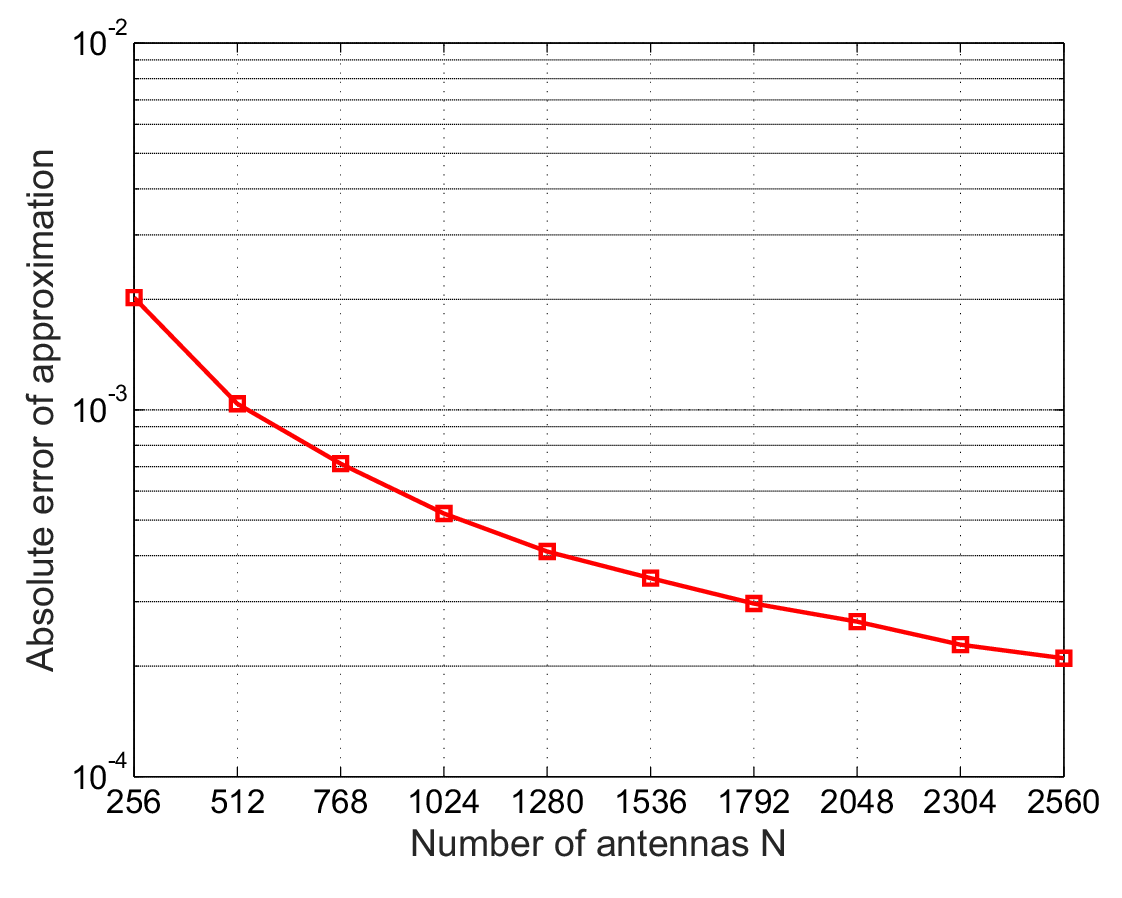} }
\caption{The absolute approximation error for the coherence of two
near-filed steering vector versus $N$} \label{f_sim_1}
\end{figure}

In Proposition~\ref{prop1_upt}, an approximate expression of the magnitude of the coherence between two near-field steering vectors
is obtained. To check the approximation accuracy of this theoretical result, we plot in
Fig.~\ref{f_sim_1} the approximation error between
the theoretical value of $M(a,b)$ and the true value of $M(a,b)$. Consider a BS with $N$
antennas operating at $100$GHz, where $N$ varies from $256$ to
$2560$. The spatial angles of the two near-field steering vectors
are randomly selected from the interval $[-1,1]$, and the
associated distance parameters are randomly sampled from the
interval $[F_r,R]$. We implement $1000$ Monte Carlo runs for
each $N$ to calculate the error between the analytical result given in (\ref{eqn11}) and the groundtruth. It can be seen that, when setting $N$ to
$256$, the approximation error of our derived analytical result is
smaller than $10^{-2}$, and such an approximation error decreases
gradually with the increase of $N$. Therefore, it is safe to conclude
that the approximation accuracy of the analytical result
(\ref{eqn11}) is accurate enough for our subsequent theoretical
analysis.

\subsection{Block-Sparsity of The Hybrid Near/Far-Field Channel}
Based on the above analysis, we now determine the condition under which $M(a,b) $ can be considered approximately equal to zero. Intuitively, $M(a,b) $ (equivalent to
$|\bm \alpha_0(m)|$) tends to zero when
$\sin(\theta_m)$ and $\sin(\theta_0)$ are
sufficiently separated. In this section, we will theoretically
analyze how far should two angles (i.e., $\sin(\theta_m)$ and
$\sin(\theta_0)$) be separated to ensure $M(a,b) $ is no
greater than a given small positive threshold $\delta$.

\subsubsection{Case of $b=0$}
We first consider the simple case where $b=0$. Under such a circumstance, we have
\begin{align}
M(a,b)  = \frac{1}{N}\left|\frac{1-e^{-jNa}}{1-e^{-ja}}\right|
\le \frac{2}{N}\left|\frac{1}{1-e^{-ja}}\right|
\end{align}
To make sure $M(a,b) <\delta$, we need
\begin{align}
\frac{2}{N}\left|\frac{1}{1-e^{-ja}}\right| < \delta
\end{align}
which gives
\begin{align}
\left|1-e^{-ja}\right| =\sqrt{2 - 2\cos(a)} > \frac{2}{N \delta}
\label{ineq_b0_0}
\end{align}
It should be noted that $\left|1-e^{-ja}\right|$ is no greater
than 2. Therefore, in~\eqref{ineq_b0_0}, we implicitly assume that
$2/(N\delta)<2$, which implies $\delta > 1/N$. Therefore, we have
\begin{align}
\cos(a) < 1 - \frac{2}{N^2\delta^2}
\end{align}
which indicates that a sufficient condition to ensure $M(a,b) <
\delta$ is given as
\begin{align}
\arccos(1-\frac{2}{N^2\delta^2}) < \left|\tilde a\right| \le \pi
\end{align}
where $\tilde a\triangleq \text{mod}(a+\pi,2\pi)-\pi$ is defined
in~\eqref{a_tilde}. Based on the relation between sufficient and
necessary conditions, we have the following result:
\begin{theorem}
\label{t_imp0}
When $\mu = \mu_0$, a necessary condition for
\begin{align}
M(a,b) \ge \delta
\end{align}
is that
\begin{align}
 \left|\tilde a \right| \le  \arccos(1-\frac{2}{N^2\delta^2})
\end{align}
\end{theorem}

According to the definition of $\tilde a$, we can rewrite the
result in Theorem~\ref{t_imp0} as
\begin{align}
&\left|\text{mod}\left(\frac{2\pi d}{\lambda}\left(\sin(\theta_m) -
\sin(\theta_0)\right) + \pi, 2\pi \right) - \pi \right | \notag\\
&\qquad \qquad\qquad\qquad\qquad\qquad\quad \le  \arccos(1-\frac{2}{N^2\delta^2})
\label{sin_n01}
\end{align}
Note that $\lambda = 2d $ and
\begin{align*}
ka\equiv \text{mod}(kb,kc) \Leftrightarrow a\equiv \text{mod}( b, c)\ \text{if} \ k \ne 0
\end{align*}
We can rewrite~\eqref{sin_n01} as
\begin{align}
&\Big|\text{mod}\left(\left(\sin(\theta_m) - \sin(\theta_0)\right) + 1, 2  \right) - 1 \Big | \le \eta_0
\label{ineq1}
\end{align}
in which $\eta_0$ is defined as
\begin{align}
\eta_0 \triangleq  \frac{1}{\pi } \arccos(1-\frac{2}{N^2\delta^2})
\end{align}

\subsubsection{Case of $b\neq 0$}
In the following, we discuss the case of $b \ne 0$. Compared with the case $b = 0$, the case of $b \ne 0$ is much more complicated since it involves the Fresnel integrals. Define $\eta_1$ and $\eta_2$ respectively as
\begin{align}
\eta_1 & \triangleq  \frac{2\sqrt{2}}{N \pi \delta} \label{def_eta1}\\
\eta_2 & \triangleq  \left(\frac{2\sqrt{2}}{N \pi \delta} + \frac{2(N-1)|b|}{\pi} \right)\notag \\
 & \overset{(a)}{=} \left(\frac{2\sqrt{2}}{N \pi \delta} + D\left|\frac{1}{\mu_0} - \frac{1}{\mu}\right| \right)
\label{def_eta2}
\end{align}
where in $(a)$ we utilized the definition of $b$. We have the following result.
\begin{theorem}
\label{the3}
When $b > 0$ (i.e., $\mu > \mu_0$), a necessary condition for $M(a,b) \ge \delta$
is
\begin{align}
-\eta_2 \le \tilde a \le \eta_1
\label{ineq20}
\end{align}
On the other hand, when $b < 0$ (i.e., $\mu < \mu_0$), this necessary condition becomes
\begin{align}
-\eta_1 \le \tilde a \le \eta_2
\label{ineq30}
\end{align}
\end{theorem}
\begin{IEEEproof}
See Appendix~\ref{app2_upt}.
\end{IEEEproof}

Similar to~\eqref{ineq1}, ~\eqref{ineq20} and~\eqref{ineq30}, respectively, can be equivalently written as
\begin{align}
-\eta_2 \le \text{mod}\left(\left(\sin(\theta_m) - \sin(\theta_0)\right) + 1, 2  \right) - 1 \le \eta_1
\label{ineq2}
\\
-\eta_1 \le \text{mod}\left(\left(\sin(\theta_m) - \sin(\theta_0)\right) + 1, 2  \right) - 1 \le \eta_2
\label{ineq3}
\end{align}

\subsubsection{Analysis}
In the above, we have derived the condition that the value of
$\tilde{a}$ (i.e., $\sin(\theta_m) - \sin(\theta_0)$) should be
satisfied to ensure $M(a,b) $ (or equivalently the $m$th element of $|\bm \alpha_0|$) is greater than a given threshold $\delta$.
Take $b < 0$ as an example. Considering the fact
$\sin(\theta_m),\sin(\theta_0)\in [-1,1]$, the
inequality~\eqref{ineq3} can be translated into the following
different cases. When $\sin(\theta_0) + \eta_2 \le 1$ and
$\sin(\theta_0) - \eta_1 \ge -1$, ~\eqref{ineq3} becomes
\begin{align}
\left\{m\Big|
\sin(\theta_0) - \eta_1 \le \sin(\theta_m) \le \sin(\theta_0) + \eta_2
\right\}
\label{cc1}
\end{align}
When $\sin(\theta_0) + \eta_2 > 1$, ~\eqref{ineq3} becomes
\begin{align}
\left\{m\Bigg|\begin{array}{l}
\sin(\theta_0) - \eta_1 \le  \sin(\theta_m) \le 1 , \\
\qquad -1 \le  \sin(\theta_m) \le -2 + \sin(\theta_0) + \eta_2 \\
\end{array}
\right\}
\label{cc2}
\end{align}
When $\sin(\theta_0) - \eta_1 < -1$, ~\eqref{ineq3} becomes
\begin{align}
\left\{m\Bigg|\begin{array}{l}
2+\sin(\theta_0) - \eta_1 \le  \sin(\theta_m) \le 1 , \\
\qquad \qquad -1 \le  \sin(\theta_m) \le \sin(\theta_0) + \eta_2 \\
\end{array}
\right\}
\label{cc3}
\end{align}

From~\eqref{cc1},~\eqref{cc2} and~\eqref{cc3}, we can see that the non-zero elements in $\bm \alpha_0$ exhibit a clustered
pattern, i.e., these non-zero entries are located either around
$\sin(\theta_0)$ (i.e.,~\eqref{cc1}) or around the endpoints of the interval (i.e.,~\eqref{cc2} or~\eqref{cc3}).

\begin{figure}[!t]
\centering {\includegraphics[width=0.36\textwidth]{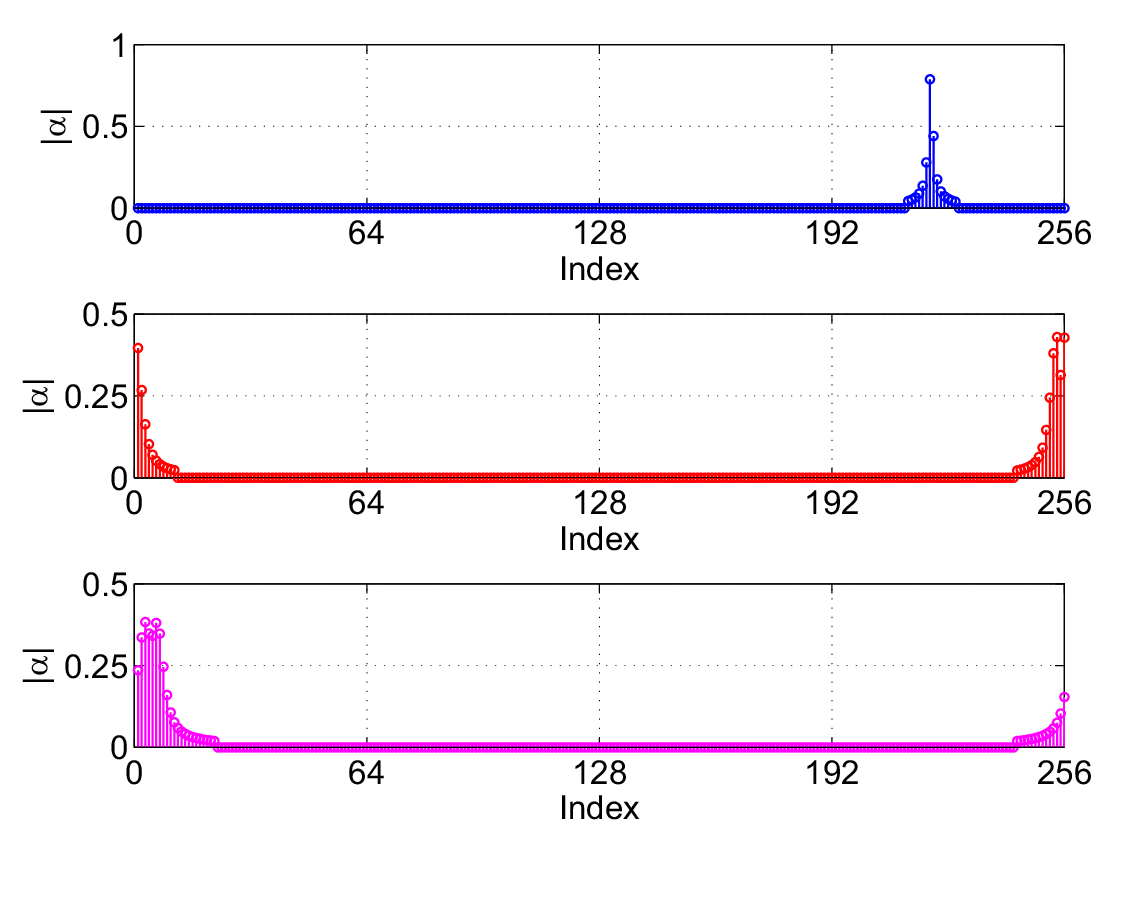} }
\caption{$|\bm\alpha_0|$ under different scenarios: $r_0 = 4,\ \theta_0 = 45^\circ$ (top); $r_0 = 4,\ \theta_0 = 75^\circ$ (middle);$r_0 = 4,\ \theta_0 = -85^\circ$ (bottom).} \label{f_pattern}
\end{figure}

To illustrate this clustered pattern more clearly, we plot the value of $|\bm \alpha_0|$ in Fig.~\ref{f_pattern} under three different scenarios when $b < 0$, where $N = 256$, $f = 100$ GHz, and $\mu = 6$. It can be observed that, depending on the value of $\sin(\theta_0)$, there are two types of block-sparsity patterns. When $\sin(\theta_0)$ is within the interval $[-1+\eta_1,1-\eta_2]$, the nonzero elements of $\bm \alpha_0$ are centered around $\sin(\theta_0)$ (see the top subfigure in Fig.~\ref{f_pattern}). Otherwise the nonzero elements of $\bm \alpha_0$ are located near the two ends of the vector $\bm \alpha_0$ (see the middle and bottom subfigures in Fig.~\ref{f_pattern}).


In the following, we would like to derive the total number of non-zero
entries in $\bm \alpha_0$, i.e., the cardinality of the set
$\{m||\bm \alpha_0(m)|\ge \delta\}$, which is denoted as $\bar K$.
From the above analysis, it can be easily verified that to ensure
$|\boldsymbol{\alpha}_0(m)|\ge \delta$, $\sin(\theta_m)$ should be within a certain interval, and such an interval can be determined via the inequalities in~\eqref{ineq1},~\eqref{ineq2}, or~\eqref{ineq3}, depending on the value of $b$. From these three inequalities, we know that the width of such an interval is given by
\begin{align}
\mathbb{L} & = \left\{
\begin{matrix}
   2\eta_0, & b = 0  \\
   \eta_1 +\eta_2, & b\ne 0 \\
   \end{matrix}
\right.
\end{align}
Note that the angular resolution of a ULA with $N$ antennas is
$2/N$, i.e., $\sin(\theta_{m+1}) - \sin(\theta_m) = 2/N$ for $m =
1,\cdots,N-1$. Therefore, $\bar K$, the cardinality of the set
$\{m||\bm \alpha_0(m)|\ge \delta\}$ can be obtained as
\begin{align}
\bar K = \left\lceil \frac{\mathbb{L}}{2/N} \right\rceil
\end{align}
which can be further simplified as
\begin{align}
\bar K & = \left\{
\begin{matrix}
   \left\lceil {\eta_0 N} \right\rceil & b = 0  \\
   \ & \ \\
   \left\lceil \frac{(\eta_1 + \eta_2) N}{2} \right\rceil & b\ne 0 \\
   \end{matrix}
\right.
\end{align}
where $\lceil \cdot \rceil$ denotes the ceiling function.

When $b = 0$, $\bar K$ is given as
\begin{align}
\bar K = \left\lceil \frac{N}{\pi} \arccos \left(1 - \frac{2}{N^2\delta^2}\right)\right\rceil
\label{barK1}
\end{align}
As $N$ tends to $\infty$, $2/(N^2\delta^2)$ tends to $0$. It is
known that the Puiseux series of $\arccos(1-x)$ around $x=0^{+}$
is given as
\begin{align}
\arccos(1-x) = \sqrt{2 x} + \frac{(2x)^{3/2}}{24} + \mathcal{O}(x^{5/2})
\end{align}
Therefore, we have
\begin{align}
\lim_{N\to \infty} \frac{N}{\pi} \arccos \left(1 - \frac{2}{N^2\delta^2}\right) = \frac{2}{\pi \delta}
\end{align}
which leads to
\begin{align}
\lim_{N\to \infty} \bar K = \left\lceil \frac{2}{\pi\delta}\right\rceil
\label{beq_103}
\end{align}
We see that when $b=0$, the number of nonzero entries in
$\boldsymbol{\alpha}_0$, $\bar K$, tends to be a constant as $N$
goes to infinity.

When $b \ne 0$, $\bar K$ is given by
\begin{align}
\bar K &= \left\lceil \frac{2\sqrt{2}}{\pi\delta }  +
\frac{ND}{2}\left|\frac{1}{\mu_0} - \frac{1}{\mu}\right| \right\rceil
\label{barK2}
\end{align}
According to Assumption 1 and the definition of $\mu$, it can be readily verified that (a detailed explanation is also
provided in ~\eqref{bound_b0}):
\begin{align}
\left|\frac{1}{\mu_0} - \frac{1}{\mu}\right| \le \frac{1}{0.62}\sqrt{\frac{\lambda}{D^3}}
\end{align}
Therefore, we have
\begin{align}
\frac{ND}{2}\left|\frac{1}{\mu_0} - \frac{1}{\mu}\right| \le \frac{N}{1.24}\sqrt{\frac{2}{N-1}}
\label{sub_N}
\end{align}
We see that when $b\neq 0$, the number of nonzero entries in
$\boldsymbol{\alpha}_0$, $\bar K$, increases sublinearly in $N$,
i.e. $\bar K\sim\sqrt{N}$. Hence the percentage of nonzero entries
in $\boldsymbol{\alpha}_0$ is at the order of $1/\sqrt{N}$, which
tends to a small value as $N$ becomes sufficiently large. Since
the non-zero elements of $\bm\alpha_{0}$ has a clustered pattern,
we can conclude that $\bm\alpha_{0}$ has a block-sparse structure. This means that
the channel $\boldsymbol{h}$ admits a block-sparse representation on the dictionary $\boldsymbol{D}_{\mu}$.

\begin{remark}
Although the above analysis focuses on the simple scenario where only the LOS path exists between the user and the BS,
our analysis can be readily extended to the general multi-path scenarios containing both LOS and NLOS components. Suppose
there are $L$ paths in total, in which case
the percentage of nonzero entries in $\boldsymbol{\beta}_{\mu}=\boldsymbol{D}_{\mu}^H\boldsymbol{h}$ is at most in the order of $L/\sqrt{N}$. Note
that due to limited scattering characteristics, $L$ is usually small for mmWave/THz channels.
In addition, the power of the LOS path is much higher (about 13 dB higher
in mmWave bands and 20dB higher in THz bands) than the combined power of the NLOS paths~\cite{AkdenizRiza14,HanWang22}.
Therefore it is expected that $\boldsymbol{\beta}_{\mu}$ still exhibits a block-sparse structure for a sufficiently large value of $N$.

The proposed dictionary in this work is an extension of the DFT matrix by
multiplying the DFT matrix with a diagonal matrix that embodies
the characteristics of the near-field property. The DFT matrix can
be considered as a special case of our dictionary with an infinitely large effective distance. Our
analysis indicates that the channel representation becomes
sparser when the effective distance associated with the dictionary
approaches the true effective distance associated with the LOS
path (cf.~\eqref{barK2}). Therefore, when the LOS path is in the near-field, our proposed dictionary is expected to yield a sparser
representation than the DFT matrix.
\end{remark}


\subsection{Numerical Validation of Theoretical Results}
In the above analysis,
we theoretically analyzed the number of nonzero elements in $\boldsymbol{\alpha}_0$,
which is given by~\eqref{barK1} or~\eqref{barK2} depending on the value of $b$.
Note that the parameter $\delta$ can be set to a small value, say, $\delta=0.01$, when
calculating the number of nonzero elements in $\boldsymbol{\alpha}_0$.

\begin{figure}[!t]
\centering {\includegraphics[width=0.41\textwidth]{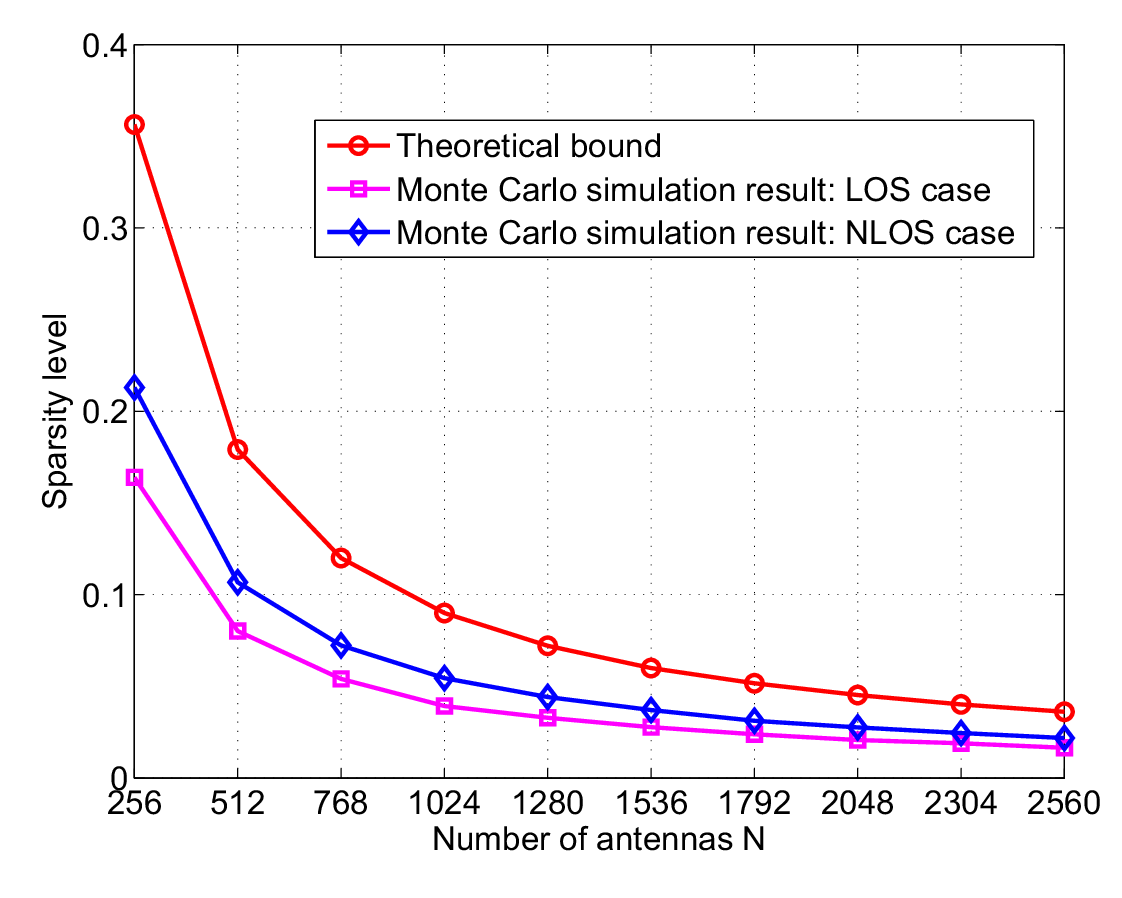} }
\caption{The sparsity level of $\bm \alpha_{0}$ versus $N$.}
\label{f_sim_2}
\end{figure}

We now provide simulations to corroborate our theoretical results. In our simulations,
the element in $\bm \alpha_{0} $ with its magnitude below $\delta=0.01$
is considered as a zero element. The sparsity level is obtained by
averaging results of $1000$ independent Monte Carlo runs. For each
Mote Carlo run, $\bm \alpha_0$ is obtained via $\bm D_{\mu}^H \bm
a(\theta_0,r_0)$, where the parameters $\{\mu ,r_0,\theta_0\}$ are
randomly generated. To corroborate the validness of our result on
multi-path channels, we also simulate the multi-path scenario containing
both LOS and NLOS components. In this case, $\bm \alpha_0$ is obtained via $\bm D_{\mu}^H (\sum_{l=0}^{(L-1)} g_0\bm a(\theta_l,r_l))$, where $L$ is set to $3$, $\sum_l|g_l|^2 = 1$, and the power ratio of the LOS component to the NLOS components is set to $13$dB. Fig.~\ref{f_sim_2}
depicts the theoretical sparsity level (i.e. the percentage of nonzero elements) result as well as the numerical results. It
can be seen that our theoretical result provides an upper
bound on the true sparsity level of $\bm \alpha_0$. The sparsity level of the multi-path scenario
is slightly larger than that of the LOS case, while both are lower than the obtained theoretical bound. In addition,
the sparsity level decreases as $N$ increases. This result corroborates our theoretical analysis that the percentage
of nonzero elements in $\bm \alpha_0$ decreases with $1/\sqrt{N}$.
Particularly, when $N$ exceeds $512$, the sparsity level
drops below $0.1$. 


\section{Channel Estimation and RIP Condition Analysis}
\label{4}
\subsection{Channel Estimation}
From the above analysis, we know that the hybrid near/far-field
channel admits a block-sparse representation on a dictionary $\bm
D_{\mu}$, i.e., $\bm h = \bm D_{\mu}\bm \beta_{\mu}$.
Substituting this sparse representation into the received signal
model~\eqref{y} results in
\begin{align}
\bm y & = \bm F \bm D_{\mu}\bm \beta_{\mu} + \bm n \triangleq \bm\Psi\bm \beta_{\mu} + \bm n
\label{y_sparse1}
\end{align}
where $ \bm\Psi\triangleq \bm F \bm D_{\mu}$ is the sensing
matrix. Note that elements of $\bm F$ can be randomly selected
from $\{+1/\sqrt{N},-1/\sqrt{N}\}$ with equal probabilities or can be
i.i.d. random variables following a complex Gaussian distribution
$\mathbb{CN}(0, {1}/{N})$.

Based on~\eqref{y_sparse1}, the hybrid near/far-filed channel
estimation problem can be converted into a block-sparse signal
recovery problem whose objective is to recover a block-sparse
$\bm\beta_{\mu}$ from noise-corrupted observations $\bm y$. Many
classical block-sparse signal recovery algorithms such as
model-based CoSaMP~\cite{BaraniukCevher10} and the
block-OMP~\cite{EldarKuppinger10} can be applied to solve our
hybrid near/far-field channel estimation problem. Note that the model-based
CoSaMP and block-OMP methods require the knowledge of
block-partition. For our considered problem, both the locations of
non-zero elements and the size of the nonzero block depend on some
unknown parameters such as $\sin(\theta_0)$. As a result, the
block partition of the block-sparse signal is usually unavailable.
In this case, the model-based CoSaMP and block-OMP methods may incur a
potential performance loss. Instead, some other block-sparse
signal recovery algorithms such as the pattern-coupled sparse Bayesian
learning~\cite{FangShen14} that do not need the knowledge of block
partition can be employed obtain a more accurate estimate of the
hyrbid near/far-field channel.

\subsection{Block-RIP Condition}
In the following, we provide an analysis to show how many
measurements are required to enable a reliable recovery of the
high-dimensional hybrid near/far channel vector. To answer this
question, we introduce the notion of block RIP which is an
extension of the RIP to a block-sparse vector. We first provide
the definition of a block $k$-sparse signal as follows:
\begin{definition}[\cite{EldarMishali09}]
Consider a vector $\bm c\in \mathbb{C}^{N}$ that can be divided
sequentially into $M$ non-overlapping blocks, i.e.,
\begin{align}
\bm c = \left[\bm c_1^H\phantom{0}\cdots\phantom{0}\bm
c_i^H\phantom{0}\cdots\phantom{0}\bm c_M^H \right]^H
\end{align}
where $\bm c_i$ is the $i$th block of length $d_i$ with $N =
\sum_{i=1}^M d_i$. Define the set
$\mathcal{I}=\{d_1,\cdots,d_M\}$. $\bm c$ is called block
$k$-sparse over $\mathcal{I}$ if
\begin{align}
\|\bm c\|_{0,\mathcal{I}} = \sum_{i=1}^M \mathbb{I}\left(\|\bm
c_i\|_{2}> 0 \right) \le k
\end{align}
where $\mathbb{I}\left(\|\bm c_i\|_{2}> 0 \right)$ is an indicator function:
\begin{align}
\mathbb{I}\left(\|\bm c_i\|_{2}> 0 \right) =
\left\{\begin{array}{ll}
1,& \|\bm c_i\|_{2}> 0\\
0,& \text{otherwise}\\
\end{array}
\right.
\end{align}
\end{definition}

With the definition of block $k$-sparse, the block RIP is defined as:
\begin{definition}[\cite{EldarMishali09}]
Consider a given matrix $\bm B\in\mathbb{C}^{n\times N}$. Then
$\bm B$ has the block RIP over the set $\mathcal{I}\triangleq
\{d_1,\cdots,d_M\}$ with parameter $\xi_{k,\mathcal{I}}$ if for
arbitrary $\bm c \in \mathbb{C}^N$ that is block $k$-sparse over
$\mathcal{I}$ we have that
\begin{align}
(1-\xi_{k,\mathcal{I}})\|\bm c\|_2^2 \le \|\bm B\bm c\|_2^2 \le (1+\xi_{k,\mathcal{I}})\|\bm c\|_2^2
\label{block_rip}
\end{align}
\end{definition}

From our previous analysis, we know that the size of the nonzero
block in $\bm \beta_{\mu}$ is in the order of $\sqrt{N}$.
Therefore we uniformly divide the $N$ entries of $\bm
\beta_{\mu}$ into $\sqrt{N}$ blocks, where we assume $\sqrt {N}$
is an integer number. Under such a circumstance, the set
$\mathcal{I}$ is written as
\begin{align}
\mathcal{I} =\{d_1,\cdots,d_M\}
\label{sparse_set}
\end{align}
with $M = \sqrt{N}$ and $d_i = \sqrt{N}, \ \forall i
\in\{1,\cdots,M\}$. According to our previous analysis, we can
conclude that $\bm \beta_{\mu}$ is block $\varrho$-sparse over
the set $\mathcal{I}$, with $\varrho$ given as
\begin{align}
\varrho & = \left\lceil  \frac{\bar K}{\sqrt{N}} \right\rceil \notag\\
& \overset{(a)}{\le}  \left\lceil \frac{2\sqrt{2}}{\pi\delta \sqrt{N}}  +
\frac{ND}{2\sqrt{N}}\left|\frac{1}{r_0} - \frac{1}{r_p}\right| \right\rceil \notag\\
&\overset{(b)}{\le}\left\lceil \frac{2\sqrt{2}}{\pi\delta \sqrt{N} }
+\frac{\sqrt{2}}{1.24}\sqrt{\frac{N}{N-1}} \right\rceil
\end{align}
where in $(a)$ we used the largest possible number of non-zero
elements in $\bm \beta_{\mu}$ and meanwhile utilized the fact that
$\left\lceil \left\lceil a \right\rceil\right\rceil \equiv
\left\lceil a \right\rceil$, and in $(b)$ we applied the inequality
in~\eqref{sub_N}. It can be readily verified that, when setting
$\delta = 0.01$ and $N \ge 256$, $\varrho$ is no greater than $7$.

For the block $\varrho$-sparse signal $\bm \beta_{\mu}$, we have
the following theorem:
\begin{theorem}
\label{blk_rip}
Let $\kappa > 0$ and $0< \xi < 1$ be constant numbers. If
\begin{align}
T \ge \frac{36}{7\xi}\left(\varrho\ln \left(\frac{e\sqrt{N}}{\varrho}\right)
+ \varrho\sqrt{N}\ln\left(\frac{12}{\xi}\right)+ \ln 2 + \kappa\right)
\end{align}
then $\bm\Psi$ satisfies the block-RIP defined
in~\eqref{block_rip} with $\xi_{\varrho,\mathcal{I}} = \xi$ with
probability at least $1-e^{-\kappa}$.
\end{theorem}
\begin{IEEEproof}
See Appendix~\ref{pf_blk_rip}.
\end{IEEEproof}

According to Theorem~\ref{blk_rip}, since $\varrho$ is generally
smaller than $\sqrt{N}$, roughly $\mathcal{O}(\sqrt{N})$
observations are required to guarantee the block-RIP of $\bm
\Psi$. In contrast, if the block-sparse structure is ignored, the
observations should be on the order of
$\mathcal{O}(\sqrt{N}\ln(N))$ to ensure $\bm \Psi$ satisfies the
standard RIP condition. Since $\sqrt{N}\ll N $ for a large value
of $N$, this means that our proposed method can achieve a
substantial training overhead reduction. On the other hand, we
would like to point out that the above analysis is pessimistic as
it applies for an arbitrary dictionary $\boldsymbol{D}_{\mu}$.
When the dictionary $\boldsymbol{D}_{\mu}$ is carefully chosen
such that $\mu$ is close to $ \mu_0$, from~\eqref{barK2} we see that
$\bar{K}$ would approach a constant, which means that the number
of nonzero entries in $\boldsymbol{\beta}_{\mu}$ remains a
constant that is independent of $N$. As a result, the number of
measurements required to recover the channel would be much
fewer than predicted by Theorem~\ref{blk_rip}.

\section{Simulation Results}
\label{5}
In this section, simulation results are presented to illustrate the performance of the proposed block-sparsity-aware channel estimation approach. We compare our method with the polar-domain based on-grid and off-grid near-field channel estimation algorithms~\cite{CuiDai22}. The transform matrix of the polar-domain method is generated according to Algorithm 1 of~\cite{CuiDai22}. In~\cite{CuiDai22}, the orthogonal matching pursuit (OMP) method
is employed to recover the sparse channel. For a fair comparison,
here we employ a block-OMP method~\cite{EldarKuppinger10} to solve
our formulated compressed sensing problem.

We consider a communication system operating at $100$GHz with a BS equipped with $N=256$ antennas. We assume that there are $L = 3$
propagation paths including a LOS path and two NLOS paths. The complex gain of the LOS path is randomly generated according to
$\mathbb{CN}(0,1)$, and the complex gains of the NLOS paths are also randomly generated such that the
ratio of the signal power of the LOS path to the power of the NLOS
paths is set to $13$dB~\cite{AkdenizRiza14}. The angle and the
distance parameters of each path are uniformly
generated according to $\theta_l \sim U[0,\pi)$ and $r_l \sim U[F_r,
1.2R]\text{m}$, where $F_r = 2.16$ and
$R = 97.54$. Therefore, the simulated channel may include both near-field and far-field components. Each element in $\bm F$ follows a complex Gaussian distribution $\mathbb{CN}(0, {1}/{N})$. The normalized mean square error (NMSE) is used as
a metric to evaluate the channel estimation performance,
which is defined as
\begin{align}
\text{NMSE}\triangleq \mathbb{E}\left({\|\bm h - \bm{\hat h}\|^2_2}/{\|\bm h \|^2_2}\right)
\end{align}
with $\bm h$ and $\bm{\hat h}$ being, respectively, the true
channel and the estimated channel. The NMSE of the channel is
obtained by averaging over $10^3$ Monte Carlo runs.

\begin{figure}[!t]
\centering
\includegraphics[width=0.40\textwidth]{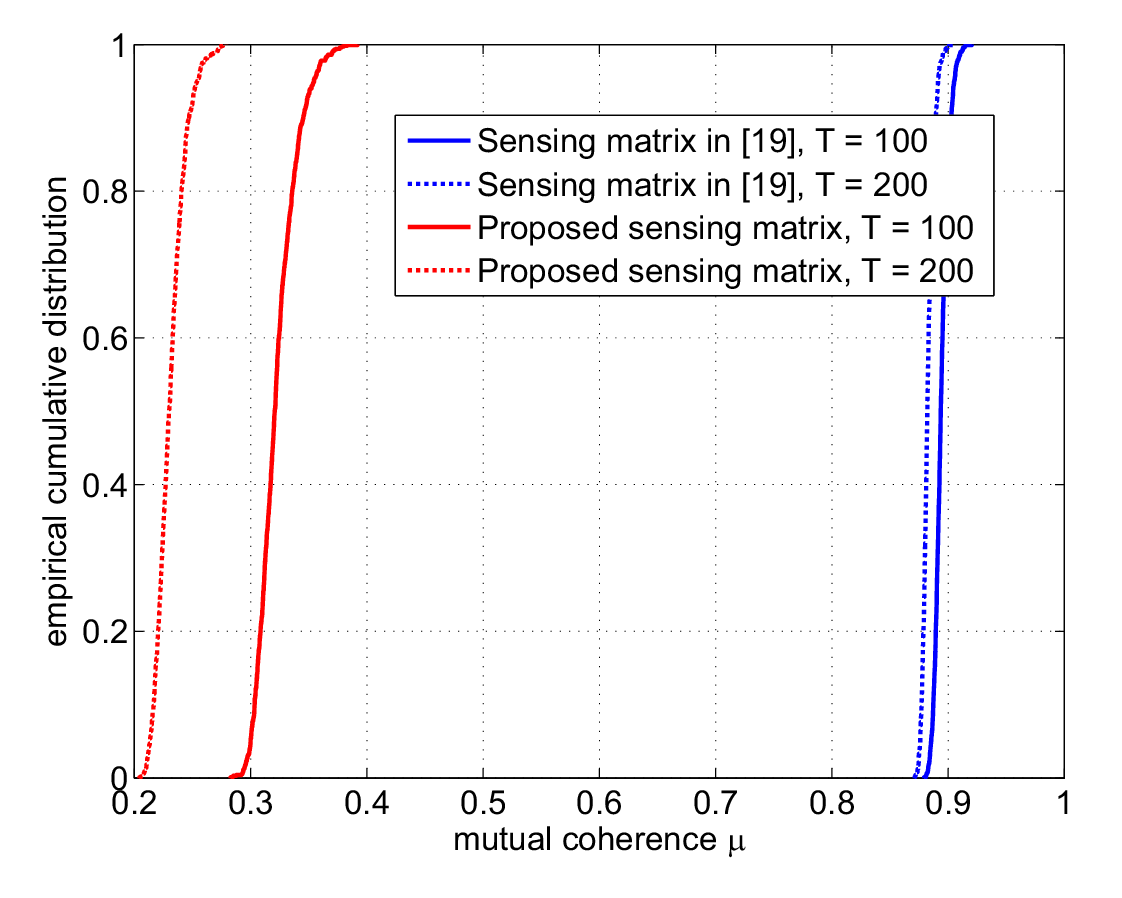}
\caption{The mutual coherence of two different sensing matrices}
\label{f_mut_cdf}
\end{figure}

\begin{figure}[!t]
\centering
\includegraphics[width=0.40\textwidth]{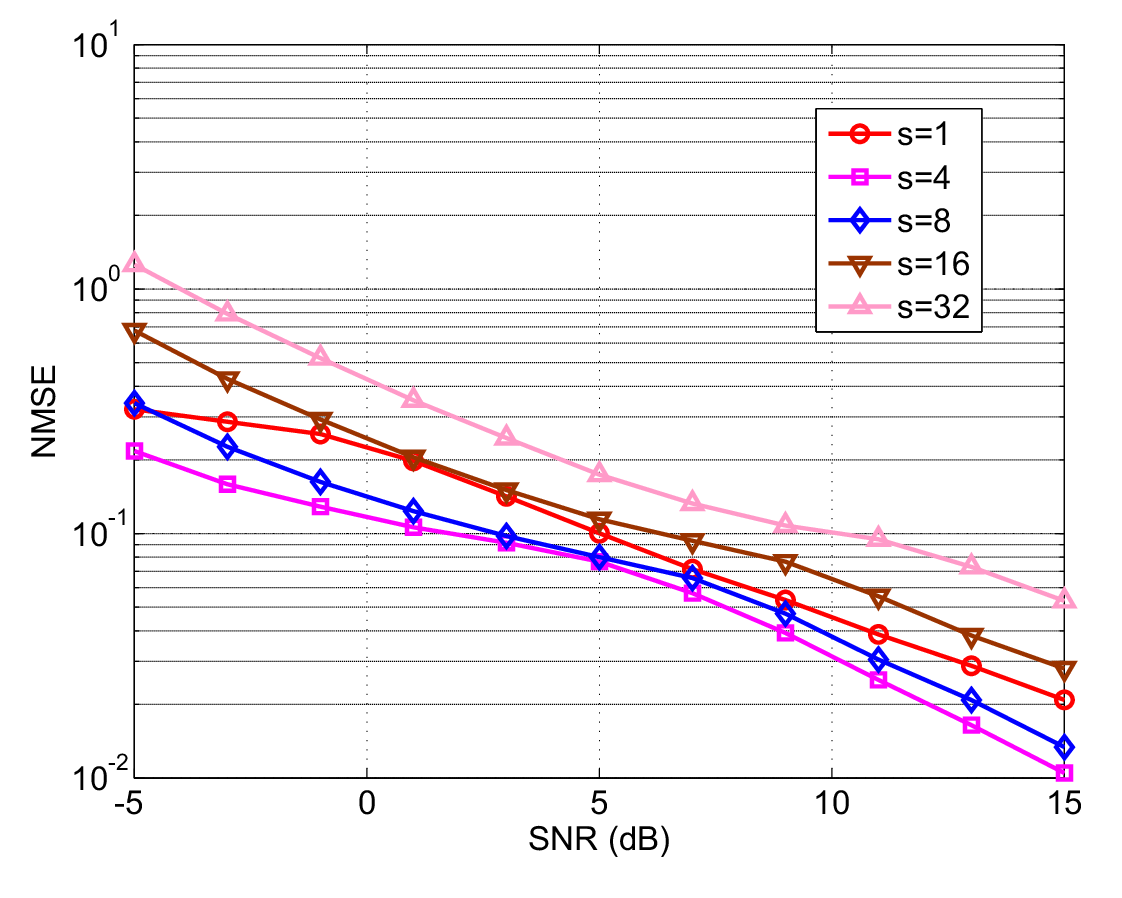}
\caption{NMSEs of different block sizes versus SNR when $T = 100$.}
\label{f_diff_blk}
\end{figure}

We first compare the mutual coherence of the sensing matrix
constructed by our method and that constructed by the polar-domain
method. The mutual coherence of a matrix is defined as the largest
coherence of any two different columns in this matrix.
Fig.~\ref{f_mut_cdf} illustrates the mutual coherence of the
sensing matrices used by the polar-domain based method and our
proposed method when $T$ is set to $100$ and $200$ respectively. These results are based on $10^3$ independent
realizations of $\bm F$. Clearly, the sensing matrix constructed
by our proposed method is more likely to have a lower mutual
coherence compared to the sensing matrix constructed by the polar-domain
based method, indicating that our sensing matrix is more favorable
for compressive sensing. This result demonstrates the superiority
of our sparse representation over the polar-domain representation.

It is known that the choice of block size (denoted by $s$ for convenience) in the block-OMP algorithm significantly impacts its signal recovery performance. Therefore, we first evaluate how the block size affects the channel estimation performance of the proposed method. In our simulations,
we set $\mu = 20$ and $T = 100$, and consider five different choices of block sizes. Fig.~\ref{f_diff_blk} shows the NMSE as a function of SNR. It can be observed that the proposed method achieves the lowest NMSE when $s=4$, followed by $s = 8$, and an excessively large block size (say $s=32$) results in notable performance loss. In the following, we set $s=4$ as a default value in our proposed method.

\begin{figure}[!t]
\centering
\subfloat[NMSEs of respective methods versus $T$ when $\text{SNR}=5$dB]
{\includegraphics[width=0.450\textwidth]{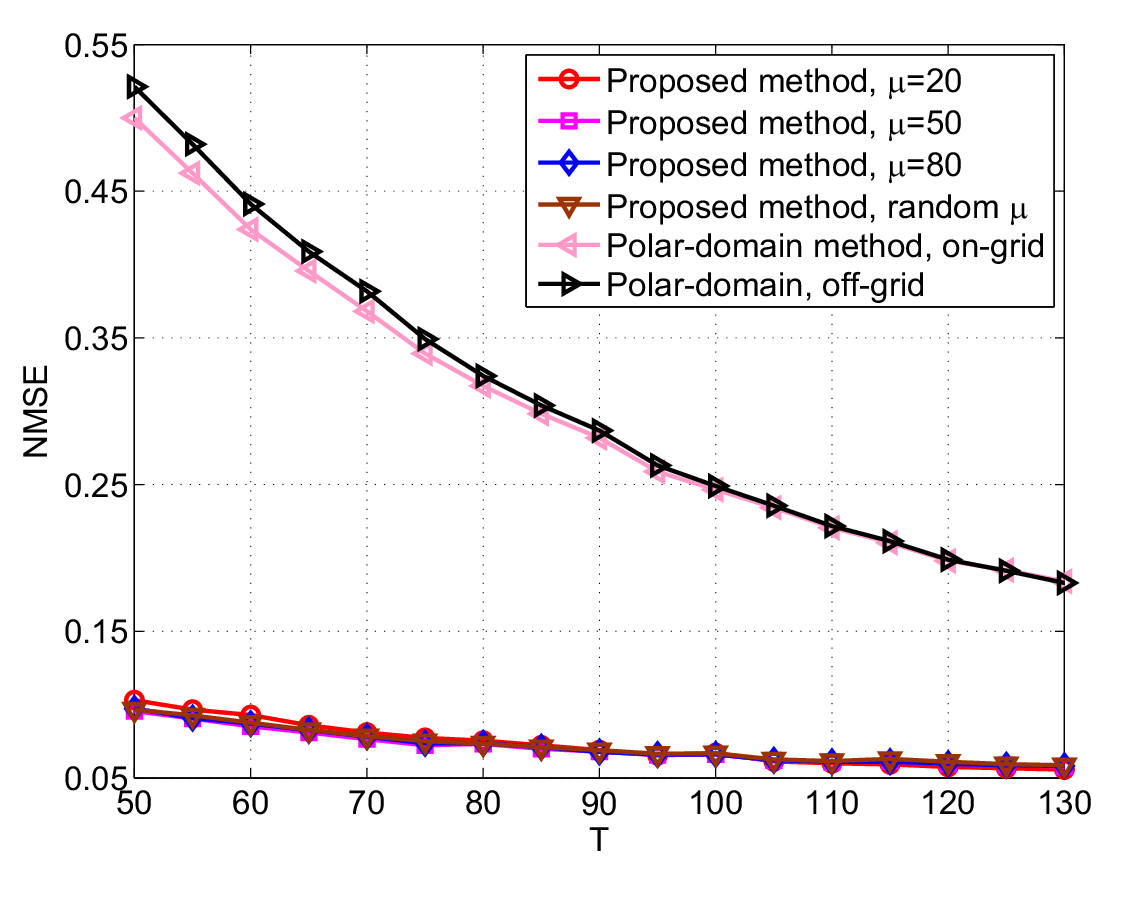}\label{fsim1}}\\
\subfloat[NMSEs of respective methods versus SNR when $T = 80$]
{\includegraphics[width=0.450\textwidth]{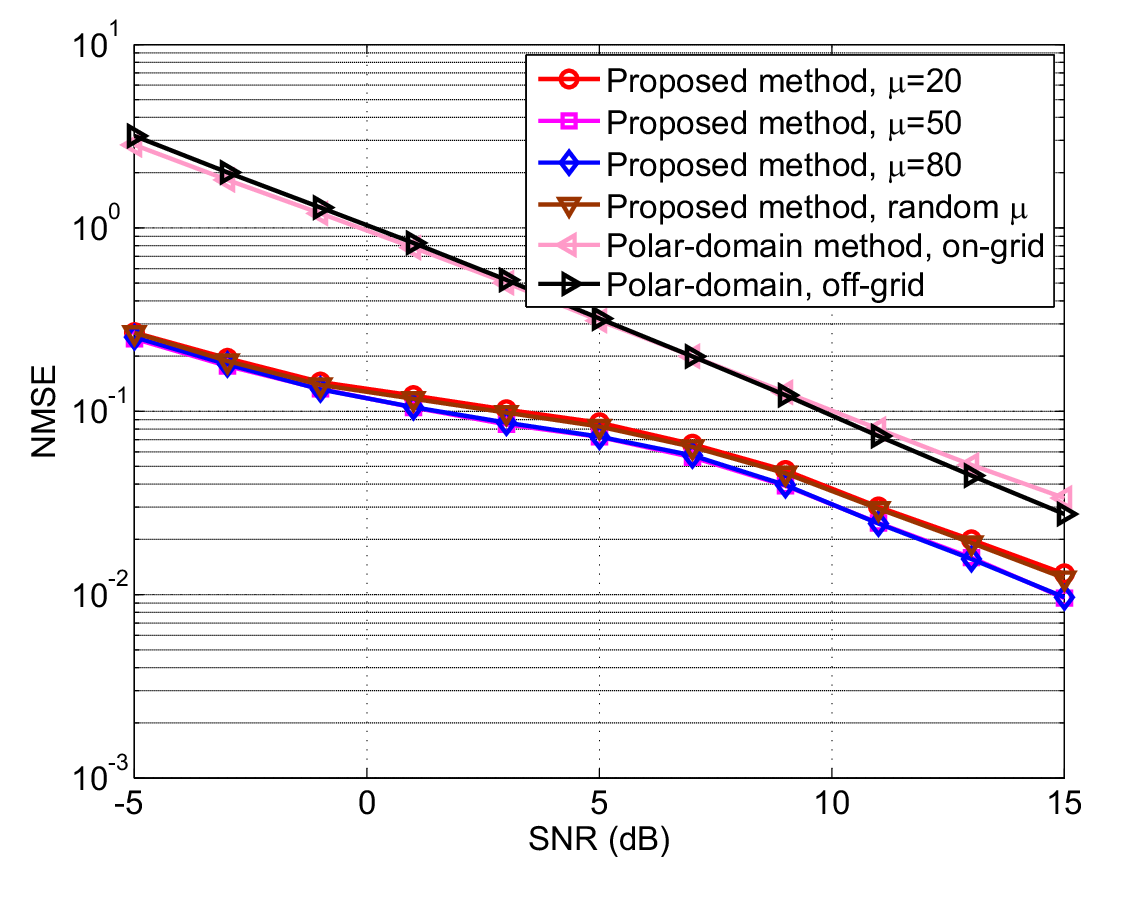}\label{fsim2}}
\caption{The NMSE of the estimated channel under different
scenarios} \label{fsim}
\end{figure}

Next, we compare our method with the polar-domain based solution. For our proposed method, we consider four different choices of the effective distance parameter $\mu$, namely, $\mu = 20$, $\mu = 50$, $\mu = 80$, and $\mu$ is randomly
selected from the interval $[F_r,R]$m. Fig.~\ref{fsim1} depicts the NMSEs as a function of the number of
measurements $T$, where the SNR is set to $5$dB. We see that the proposed method presents a clear performance advantage over
the polar-domain on-grid and off-gird channel estimation methods.
The performance gap becomes more pronounced when the number of
measurements $T$ is small. Fig.~\ref{fsim2} shows NMSEs versus
SNR, with $T$ set to $T=80$. From Fig.~\ref{fsim2}, we see that our method provides an estimation error lower than $0.1$ when the SNR
is above $1$dB, while the polar-domain method requires a higher
SNR level (around $10$ dB) to achieve the same performance. This result,
again, demonstrates the superiority of our sparse representation of the hybrid near/far-field channel over the polar-domain representation. Also,
our proposed method yields similar NMSE results for different choices of $\mu$. Here the NMSE results
are averaged over $10^3$ randomly generated channels. Hence the estimation result does not favor any specific choice of $\mu$.

\begin{figure}[!t]
\centering
{\includegraphics[width=0.450\textwidth]{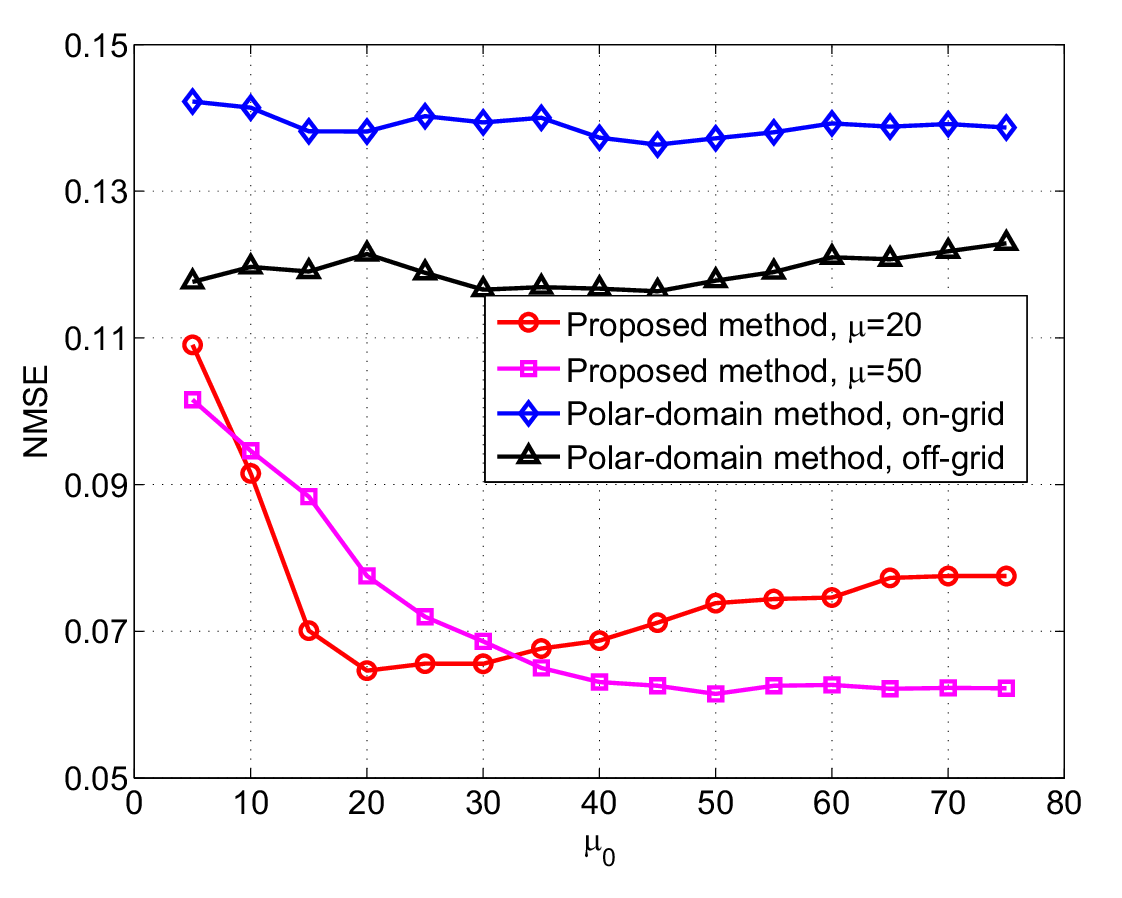}}\\
\caption{The NMSEs of the estimated channel against $\mu_0$ when $T=100$ and SNR=$5$ dB}
\label{fsim_los}
\end{figure}

The NMSE performance with respect to $\mu_0$ (i.e., the effective distance of the strongest prorogation path) is also evaluated, as illustrated in Fig.~\ref{fsim_los}. $\mu_0$ is generated by randomly selecting $r_0$ and $\theta_0$. It can be observed in Fig.~\ref{fsim_los} that the NMSEs of the polar-domain based method, whether in the on-grid or the off-grid setup, remain nearly constant. In contrast, the performance of the proposed method varies with $\mu_0$. The proposed method with the choice of $\mu = 20$ achieves the lowest NMSE when $\mu_0$ is $20$, and the same holds true for $\mu=50$. This is because, when $\mu=\mu_0$, our proposed dictionary $\boldsymbol{D}_{\mu}$ yields the sparest channel representation, which in turn leads to better estimation performance.

\section{Conclusions}
\label{6}
We investigated the hybrid near/far-field channel estimation
problem for mmWave/THz systems with an ELAA. We proposed a novel
sparse representation for hybrid near/far-field channels by
introducing a specially designed unitary matrix.
Through theoretical analysis, we discovered that the hybrid
near/far-field channel admits a block-sparse structure on the
constructed unitary matrix. By leveraging this block sparsity, the
hybrid near/far-field channel estimation problem can be converted
into a block-sparse signal recovery problem. Simulation results demonstrated that our
proposed method presents a clear performance advantage over
state-of-the-art hybrid near/far channel estimation methods that
are developed based on the polar domain representation.


\appendices

\section{Proof of Proposition~\ref{prop1_upt}}
\label{prop1_upt_proof}

We first consider the case with $b=0$. In this case, $f(a,b)$ degenerates into the coherence of two far-field steering vectors, and
$M(a,b)$ can be obtained as
\begin{align}
M(a,b) & = \frac{1}{N}\left|\sum_{n=1}^Ne^{-j  (n-1) a }\right|\notag\\
& =\frac{1}{N}\left|\frac{1-e^{-jNa}}{1-e^{-ja}}\right|
\label{eq}
\end{align}

The case with $b\ne 0$ is more complicated. For convenience, we first consider the case where $b > 0$. Recalling~\eqref{Lambda-def}, we have
\begin{align}
f(a,b) = \frac{1}{N} \sum_{n=0}^{N-1} e^{-ja n }e^{-j b n^2}
\end{align}
$f(a,b)$ is also known as the {\emph{generalized quadratic Gauss
sums}}~\cite{BerndtEvans98}. Due to the periodicity of the complex-exponential function, we first discuss the range of the two variables $a$ and $b$. Let $d=\lambda/2$, we have
\begin{align}
a = \pi (\sin(\theta_m) - \sin(\theta_0))
\end{align}
which indicates $a\in [-2\pi,2\pi]$. The value of $b$ can be
written as
\begin{align}
b = \frac{\pi \lambda}{4}\left(\frac{1}{\mu_0} - \frac{1}{\mu }\right)
\label{def_b}
\end{align}
According to Assumption 1
$r_0 \ge F_r \triangleq  0.62 \sqrt{\frac{D^3}{\lambda}}$,
we have
\begin{align}
\frac{1}{\mu_0} \triangleq \frac{\cos^2(\theta_0)}{r_0}\le \frac{1}{r_0}\le \frac{1}{0.62}\sqrt{\frac{\lambda}{D^3}}
\label{mu0}
\end{align}
In addition, $\mu$ is the user-defined parameter, and hence it can be selected as
\begin{align}
\frac{1}{\mu} \le \frac{1}{0.62}\sqrt{\frac{\lambda}{D^3}}
\label{mu}
\end{align}
Combining~\eqref{mu0} and~\eqref{mu}, we have
\begin{align}
\left|\frac{1}{\mu_0} - \frac{1}{\mu}\right| \le \frac{1}{0.62}\sqrt{\frac{\lambda}{D^3}}
\label{bound_b0}
\end{align}
Utilizing the relationship~\eqref{bound_b0}, the bound of $|b|$ is given by
\begin{align}
|b| &\le \frac{\pi \lambda}{4 } \frac{1}{0.62}\sqrt{\frac{\lambda}{D^3}}\notag\\
& = \frac{\pi}{4\times 0.62} \sqrt{\frac{\lambda^3}{D^3}}\notag\\
& \overset{(a)}{=} \frac{\pi}{4\times 0.62} \sqrt{\frac{\lambda^3}{(N-1)^3d^3}}\notag\\
& \overset{(b)}{<} \frac{\pi}{1.24(N-1)}\sqrt{\frac{2}{N-1}}
\overset{(c)}{<} \pi \label{cond_b2}
\end{align}
where $(a)$ is due to $D=(N-1)d$, $(b)$ follows from
$d=\lambda/2$, and $(c)$ holds true when $N >2$.

Since $a\in [-2\pi,2\pi]$, to guarantee the uniqueness of $f(a,b)$, we define $\tilde a$ as
\begin{align}
\tilde a \triangleq \text{mod}(a+\pi,2\pi) - \pi,
\end{align}
such that $\tilde a\in [-\pi,\pi]$. Meanwhile, we have
\begin{align}
e^{-ja n } = e^{-j \tilde a n }
\end{align}
due to the fact that $n\in\{0,\cdots, N-1\}$. Therefore, we have the following result:
\begin{align}
f(a,b) & = \frac{1}{N} \sum_{n=0}^{N-1} e^{-ja n }e^{-j b n^2}\notag\\
& = \frac{1}{N} \sum_{n=0}^{N-1} e^{-j\tilde a n }e^{-j b n^2} \triangleq f(\tilde a ,b )
\label{Eq1}
\end{align}

In addition, $f(\tilde a ,b )$ can be approximated as
\begin{align}
f(\tilde a ,b ) & =\frac{1}{N} \sum_{n=0}^{N-1} e^{-j\tilde a n }e^{-j b n^2}\notag\\
&\overset{(a)}{\approx} \frac{1}{N}\int_{n=0}^{N-1} e^{-j\tilde a n }e^{-j b n^2} dn \notag\\
& \overset{(b)}{=}\frac{1}{N} \int_{n=0}^{N-1} e^{-j (\sqrt{b} n + \frac{ \tilde a}{2\sqrt{b}})^2 -
j\frac{ \tilde a^2}{4b}} dn  \notag\\
& = \frac{1}{N} e^{- j\frac{ \tilde a^2}{4b}}\int_{n=0}^{N-1} e^{-j (\sqrt{b} n
+ \frac{ \tilde a}{2\sqrt{b}})^2 } dn \notag\\
& = \frac{1}{N} \frac{e^{- j\frac{ \tilde a^2}{4b}}}{\sqrt{b}}\int_{n=
\frac{\tilde a}{2\sqrt{b}}}^{(N-1)\sqrt{b} + \frac{ \tilde a}{2\sqrt{b}}} e^{-j n^2 } dn \notag\\
& = \frac{1}{N} \frac{e^{- j\frac{ \tilde a^2}{4b}}}{\sqrt{b}}\left( f_1(a,b) - j f_2(a,b)\right)
\label{Eq2}
\end{align}
where in $(a)$, the summation is approximated by an integral~\cite{Paris08}, and
$(b)$ is due to the assumption $b>0$, $f_1(a,b)$ and $f_2(a,b)$
are, respectively, defined in~\eqref{f_1} and~\eqref{f_2}.

Combining~\eqref{Eq1} and~\eqref{Eq2}, we have
\begin{align}
M(a,b) \approx\frac{1}{N\sqrt{|b|}} \sqrt{|f_1(a,b)|^2 + |f_2(a,b)|^2}
\label{f_ab}
\end{align}
Thus we obtain the expression of $M(a,b)$ for the case $b
>0$.

When $b < 0$, the expression in~\eqref{f_ab} still holds true,
except that the Fresnel integral should be generalized to the
complex domain. We omit the details due to the similarities.

This completes our proof.

\section{Proof of Theorem~\ref{the3}}
\label{app2_upt}

We first consider the case of $b>0$. In this case, the result depends on the sign of $\tilde a$.
\subsection{When $\tilde a >0$}
Before discussing the case where $b>0$ and $\tilde a >0$, we first
introduce the following proposition:
\begin{proposition}
\label{prop2}
If $x >0$ and $\Delta > 0$, then the following inequalities hold:
\begin{align}
\left|C(x+\Delta) - C( x) \right| &< \frac{1}{x}\\
\left|S(x+\Delta) - S( x) \right| &< \frac{1}{x}\label{ss2}
\end{align}
\end{proposition}
\begin{IEEEproof}
See Appendix~\ref{app2}.
\end{IEEEproof}

Applying Proposition~\ref{prop2} to $f_1(a,b)$ and $f_2(a,b)$, we
have
\begin{align}
|f_1(a,b)|< \frac{2\sqrt{b}}{ \tilde a},\ |f_2(a,b)|<\frac{2\sqrt{b}}{ \tilde a}
\label{c1}
\end{align}
As a result, we have
\begin{align}
M(a,b) \!\! =\!\! \frac{1}{N\sqrt{b}} \sqrt{f_1^2(a,b) + f_2^2(a,b)}\!\! < \!\!
\frac{2\sqrt{2} }{N  \tilde a}
\end{align}
To ensure $M(a,b)< \delta$, the following inequality
should be satisfied
\begin{align}
\frac{2\sqrt{2} }{N  \tilde a} <\delta
\end{align}
which leads to
\begin{align}
\frac{2\sqrt{2}}{N \delta} <\tilde a \le \pi
\label{ine_a}
\end{align}
Note that in~\eqref{ine_a} we implicitly assume that
\begin{align*}
\frac{2\sqrt{2}}{N \delta} <\pi
\end{align*}
which is equivalent to
\begin{align}
\delta > \frac{2\sqrt{2}}{N \pi} \approx \frac{0.9}{N}
\end{align}
The above condition can be met for a sufficiently large $N$.

\subsection{When $\tilde a < 0$}
The case with $b>0$ and $\tilde a  < 0$ is much more complex since
we cannot directly utilize Proposition~\ref{prop2}. To deal with
this challenge, we apply the following properties of the Fresnel
integrals, i.e., $C(-\zeta)= -C(\zeta)$ and $S(-\zeta)=-S(\zeta)$,
and rewrite $f_1(a,b)$ and $f_2(a,b)$ as follows
\begin{align}
f_1(a,b) &\triangleq C\left(\frac{-\tilde a}{2\sqrt{b}}\right) -
C\left( \frac{ -\tilde a}{2\sqrt{b}} - (N-1)\sqrt{b}\right)\\
f_2(a,b) &\triangleq S\left(\frac{-\tilde a}{2\sqrt{b}}\right) -
S\left( \frac{-\tilde a}{2\sqrt{b}} - (N-1)\sqrt{b}\right)
\end{align}

If $- {\tilde a}/{(2\sqrt{b})} - (N-1)\sqrt{b} > 0$, we have, according to Proposition~\ref{prop2},
\begin{align}
|f_1(a,b)|&< \frac{1}{- \frac{\tilde a}{2\sqrt{b}} - (N-1)\sqrt{b}}\\ |f_2(a,b)|&
<\frac{1}{- \frac{\tilde a}{2\sqrt{b}} - (N-1)\sqrt{b}}
\end{align}
which gives
\begin{align}
M(a,b) < \frac{\sqrt{2}}{N\sqrt{b}} \frac{1}{- \frac{\tilde a}{2\sqrt{b}} - (N-1)\sqrt{b}}
\end{align}
Let
\begin{align}
\frac{\sqrt{2}}{N\sqrt{b}} \frac{1}{- \frac{\tilde a}{2\sqrt{b}} - (N-1)\sqrt{b}} <\delta
\end{align}
We have
\begin{align}
-\pi \le \tilde a  < -\left(\frac{2\sqrt{2}}{N \delta} + 2(N-1)b \right)
\label{a_tilde_2}
\end{align}

On the other hand, when $- {\tilde a}/{(2\sqrt{b})} - (N-1)\sqrt{b} \le  0$, we have
\begin{align}
f_1(a,b) &= C\left(\frac{-\tilde a}{2\sqrt{b}}\right) + C\left(\frac{\tilde a}{2\sqrt{b}} + (N-1)\sqrt{b}\right)\\
f_2(a,b) &= S\left(\frac{-\tilde a}{2\sqrt{b}}\right) + S\left(
\frac{\tilde a}{2\sqrt{b}} + (N-1)\sqrt{b}\right)
\end{align}
Based on the fact $C(\zeta) < 0.98$ and $S(\zeta) < 0.90$ for
$\zeta > 0$ (the validness of the results can be numerically verified), we can arrive at
\begin{align}
M(a,b) = \frac{1}{N\sqrt{b}}\sqrt{f_1^2(a,b) + f_2^2(a,b)} < \frac{2.7}{N\sqrt{b}}
\end{align}
Thus, we can let
\begin{align}
\sqrt{b} > \frac{2.7}{N\delta}
\label{cond_b}
\end{align}
to ensure $M(a,b) <\delta$. Meanwhile, from~\eqref{cond_b2} we
know that
\begin{align}
b \le \frac{\pi}{1.24(N-1)}\sqrt{\frac{2}{N-1}}
\label{cond_b3}
\end{align}
To guarantee~\eqref{cond_b} and~\eqref{cond_b3} hold
simultaneously, the following condition should be satisfied
\begin{align}
\frac{(N\delta)^2}{(N-1)\sqrt{N-1}} > 2.035
\end{align}
Nevertheless, for a large value of $N$ and a small positive
$\delta$, the above condition does not hold valid. Therefore, we
can draw a conclusion that when $- {\tilde a}/{(2\sqrt{b})} -
(N-1)\sqrt{b} \le 0$, there is no way to ensure $|f(a,b)|$ is
smaller than $\delta$.

Based on the above analysis, we can conclude that when $b>0$,
$\tilde a$ needs to satisfy the following condition in order to
guarantee $M(a,b) < \delta$:
\begin{align}
-\pi \le \tilde a  < -\left(\frac{2\sqrt{2}}{N \delta} + 2(N-1)b \right)
\ \text{or}\ \frac{2\sqrt{2}}{N \delta} < \tilde a \le \pi
\end{align}
According to the relation between sufficient conditions and
necessary conditions, we complete the proof of the first part of Theorem~\ref{the3}.

\subsection{The Case of $b < 0$}
When $b < 0$, i.e., $\mu <
\mu_0$, using the definition of $M(a,b)$ we have $M(a,b) = M(-a,-b)$. By utilizing this relationship and the first part of Theorem~\ref{the3}, we can arrive at the result of the second part of Theorem~\ref{the3}. This completes our
proof.

\section{Proof of Proposition~\ref{prop2}}
\label{app2}

Based on the definition of $C(x)$, we have
\begin{align}
&C(x+\Delta)  - C( x) \notag\\
&\qquad= \int_{x}^{x+\Delta}\cos(t^2)dt\notag\\
&\qquad=\int_{x}^{x+\Delta}\frac{1}{2t}d\left(\sin(t^2)\right)\notag\\
&\qquad \overset{(a)}{=} \left.\frac{\sin(t^2)}{2t}\right|_{t = x}^{t =
x+ \Delta} + \int_{x}^{x+\Delta}\frac{\sin(t^2)}{2t^2}dt
\end{align}
where in $(a)$ we utilize the integration by parts. Therefore, $\left|C(x+\Delta)  - C( x) \right|$
can be expressed by
\begin{align}
&\left|C(x+\Delta) - C( x) \right|\notag\\
& \le \left| \frac{\sin((x+\Delta)^2)}{2(x+\Delta)} \right| + \left| \frac{\sin(x^2)}{2x} \right|
+ \int_{x}^{x+\Delta} \left|\frac{\sin(t^2)}{2t^2}\right|dt\notag\\
& \le \frac{1}{2(x+\Delta)} + \frac{1}{2x} +  \int_{x}^{x+\Delta}  \frac{1}{2t^2}dt = \frac{1}{x}
\end{align}
The case for~\eqref{ss2} can be similarly verified. This completes the proof.

\section{Proof of Theorem~\ref{blk_rip}}
\label{pf_blk_rip} Before proving this theorem, we have the
following propositions:
\begin{proposition}
\label{pp_1} Each element in $\bm\Psi\in\mathbb{C}^{T\times N}$
follows $\mathbb{CN}(0, {1}/{N})$ with overwhelming probability.
Also entries of $\bm\Psi$ are mutually independent.
\end{proposition}
\begin{IEEEproof}
See Appendix~\ref{prop_1}.
\end{IEEEproof}

\begin{proposition}[\cite{EldarMishali09,BlumensathDavies09}]
\label{pp_2} Consider a random Gaussian matrix $\bm B$ of size
$n\times N$ and block $k$-sparse signals over $\mathcal{I} = \{d_1 =
d,\cdots, d_M = d\}$, where $N = Md$ for some integer $M$. Let
$\kappa>0$ and $0<\xi<1$ be constant numbers. If
\begin{align}
n \ge \frac{36}{7\xi}\left(\ln(2L) + kd\ln\left(\frac{12}{\xi}\right)+\kappa\right)
\end{align}
where $L =\begin{pmatrix} M\\k\end{pmatrix}$, then $\bm B$
satisfies the block-RIP~\eqref{block_rip} with
restricted isometry constant $\xi_{k,\mathcal{I}} = \xi$,
with probability at least $1-e^{-\kappa}$.
\end{proposition}

It is known that $\bm \alpha_{\mu}$ is block $\varrho$-sparse
vector over $\mathcal{I}$ defined in~\eqref{sparse_set}. Directly
applying Proposition~\ref{pp_1} and Proposition~\ref{pp_2} will
lead to the result that if the number of observations $T$
satisfies
\begin{align}
T \ge \frac{36}{7\xi}\left(\ln\begin{pmatrix} \sqrt{N}\\ \varrho\end{pmatrix}
+ \varrho\sqrt{N}\ln\left(\frac{12}{\xi}\right) + \ln 2+\kappa\right)
\label{t_cond_1}
\end{align}
then $\bm \Psi$ satisfies the block-RIP~\eqref{block_rip} with restricted isometry constant
$\xi_{\varrho,\mathcal{I}} = \xi$, with probability at least
$1-e^{-\kappa}$. In addition, based on the Stirling's formula, we have
\begin{align}
\begin{pmatrix} \sqrt{N}\\\varrho\end{pmatrix} \le \left(\frac{e\sqrt{N}}{\varrho}\right)^{\varrho}
\label{t_cond_2}
\end{align}
Combining~\eqref{t_cond_1} and~\eqref{t_cond_2} together will complete our proof.

\section{Proof of Proposition~\ref{pp_1}}
\label{prop_1} The $(i,j)$th element of $\bm\Psi$, denoted by
$\psi_{ij}$ is defined by
\begin{align}
\psi_{ij}& = \bm F(i,:)\bm D_{\mu}(:,j) \notag\\
&= \sum_{n=1}^N \bm F(i,n)\bm D_{\mu}(n,j)  \triangleq \sum_{n=1}^N z_n
\end{align}
where $z_n \triangleq \bm F(i,n)\bm D_{\mu}(n,j)$. Apparently, we have
\begin{align}
\mathbb{E}(z_n) & = 0\\
\mathbb{V}(z_n) & = {|\bm D_{\mu}(n,j)|^2}/{N}
\end{align}
In addition, due to the fact that $\bm F(i,n_1)$ and $\bm
F(i,n_2)$ are independent, $z_{n_1}$ and $z_{n_2}$ are
independent. Therefore, $\psi_{ij}$ is a sum of i.i.d zero-mean
random variables $\{z_n\}_{n=1}^N$. According to the central limit
theorem (CLT), we have
\begin{align}
\psi_{ij}\overset{(c)}{\sim}\mathbb{CN}\left(0,\left({\sum_{n=1}^N {|\bm D_{\mu}(n,j)|^2}}\right)/{N}\right)
\end{align}
where $\overset{(c)}{\sim}$ means ``converges to a distribution''.

Furthermore, consider two different elements in $\bm \Psi$, i.e.,
$\psi_{i_1j_1}$ and $\psi_{i_2j_2}$ with $i_1 \ne i_2$ and
meanwhile $j_1 \ne j_2$, which are expressed as
\begin{align}
\psi_{i_ij_1}& = \bm F(i_1,:)\bm D_{\mu}(:,j_1)\\
\psi_{i_2j_2}& = \bm F(i_2,:)\bm D_{\mu}(:,j_2)
\end{align}
Due to the fact that $\bm F(i_1,:)$ and $\bm F(i_2,:)$ are
independent and meanwhile $\bm D_{\mu}(:,j_1)$ and $\bm
D_{\mu}(:,j_2)$ are mutual orthogonal, $\psi_{i_1j_1}$ and
$\psi_{i_2j_2}$ are mutually independent. This can be further
identified via checking the mutual information of $\psi_{i_1j_1}$
and $\psi_{i_2j_2}$. This ends our proof.

\end{document}